\newif\ifreviewtemplate
\newif\ifreviewcolor

\reviewtemplatefalse

\reviewcolortrue

\ifreviewtemplate
    \documentclass[authoryear,preprint,review,12pt]{elsarticle}
\else
    \documentclass[final,1p,times,authoryear]{elsarticle}
   
\fi

\usepackage{amssymb}
\usepackage{amsmath}

\usepackage{lineno}

\journal{Computer Speech \& Language}

\usepackage{amsmath}
\usepackage{amsfonts}
\usepackage{array}
\usepackage{mathtools}
\usepackage{amssymb}
\usepackage{booktabs}
\usepackage{changepage}
\usepackage{tipa}

\usepackage{graphicx}
\usepackage{xspace}
\usepackage{siunitx}
\usepackage{mathabx}
\usepackage{xcolor}
\usepackage{colortbl}		
\usepackage{subcaption}
\usepackage{hyperref}
\usepackage{cleveref}

\ifreviewcolor
    
\else
    
\fi

\definecolor{gipsaPurple}{rgb}{0.34, 0.18, 0.35}
\definecolor{gipsaBlue}{rgb}{0, 0.34, 0.61}
\definecolor{gipsaDark}{rgb}{0.098, 0.133, 0.220}
\definecolor{gipsaDPC}{rgb}{0.64, 0.70, 0.14}
\definecolor{gipsaDAUTO}{rgb}{0.68, 0.20, 0.38}
\definecolor{gipsaPAD}{rgb}{0.745, 0.0, 0.0}
\definecolor{gipsaPSD}{rgb}{0.922, 0.49, 0.137}
\definecolor{gipsaPPC}{rgb}{0.0, 0.686, 0.314}
\definecolor{gipsaGAIA}{rgb}{0.0, 0.686, 0.941}
\definecolor{darkgreen}{rgb}{0.0, 0.4, 0.2}
\definecolor{darkorange}{rgb}{0.900, 0.800, 0.0}
\definecolor{lightred}{rgb}{0.950, 0.0, 0.10}

\newcommand{\colpurple}{\color{gipsaPurple}}

\newcommand{\colgreen}{\color{gipsaPPC}}

\newcommand{\colorange}{\color{gipsaPSD}}

\newcommand{\colred}{\color{gipsaPAD}}

\newcommand{\collightblue}{\color{gipsaGAIA}}

\newcommand{\M}{\mathtt{PT}}
\newcommand{\I}{\mathtt{FT}}
\newcommand{\inpaintM}{$\mathcal{I}_{\M}$\xspace}
\newcommand{\inpaintI}{$\mathcal{I}_{\I}$\xspace}
\newcommand{\inpaintT}{$\mathcal{I}_{\mathtt{ASR}}$\xspace}
\newcommand{\inpaintLI}{$\mathcal{I}_{\mathtt{LI}}$\xspace}
\newcommand{\inpaint}{$\mathcal{I}$\texttt{}\xspace}

\newcommand{\methods}{\emph{frameworks}\xspace}
\newcommand{\symbSize}{\large}
\newcommand{\tabSize}{\scriptsize}

\newcommand{\symbsigcircle}[1]{\symbSize$#1{\bullet}$\tabSize}
\newcommand{\symbsigsquare}[1]{\symbSize$#1{\sqbullet}$\tabSize}
\newcommand{\symbsigdiamond}[1]{\symbSize$#1{\blackdiamond}$\tabSize}
\newcommand{\symbsigstar}[1]{\symbSize$#1{\star}$\tabSize}
\newcommand{\symbsigtriangled}[1]{\symbSize$#1{\blacktriangledown}$\tabSize}
\newcommand{\symbsigtriangleu}[1]{\symbSize$#1{\blacktriangleup}$\tabSize}
\newcommand{\symbsigtriangler}[1]{\symbSize$#1{\blacktriangleright}$\tabSize}
\newcommand{\symbsigtrianglel}[1]{\symbSize$#1{\blacktriangleleft}$\tabSize}

\crefname{figure}{Fig.}{Figs.}

\begin{document}
\begin{frontmatter}



\title{Is Self-Supervised Learning Enough to Fill in the Gap?\\A Study on Speech Inpainting}

\author[gipsa,jena]{Ihab Asaad\fnref{fn1}}
\author[gipsa,vogo]{Maxime Jacquelin\fnref{fn1}}
\author[gipsa]{Olivier Perrotin}
\author[gipsa]{Laurent Girin}
\author[gipsa]{Thomas Hueber\corref{cor1}}

\cortext[cor1]{Corresponding author}
\fntext[fn1]{Ihab Asaad and Maxime Jacquelin contributed equally to this work.}

\affiliation[gipsa]{organization={Univ. Grenoble Alpes, CNRS, Grenoble-INP, GIPSA-lab},
            city={Grenoble},
            postcode={F-38000},
            country={France}}
\affiliation[jena]{organization={Friedrich Schiller Universität Jena},
            city={Jena},
            country={Germany}}

\affiliation[vogo]{organization={Vogo},
            city={Bernin},
            postcode={38190},
            country={France}}

\begin{abstract}
Speech inpainting consists in reconstructing corrupted or missing speech segments using surrounding context, a process that closely resembles the pretext tasks in Self-Supervised Learning (SSL) for speech encoders. 
This study investigates using SSL-trained speech encoders for inpainting without any additional training beyond the initial pretext task, and simply adding a decoder to generate a waveform. We compare this approach to supervised fine-tuning of speech encoders for a downstream task---here, inpainting. Practically, we integrate HuBERT as the SSL encoder and HiFi-GAN as the decoder in two configurations: (1) fine-tuning the decoder to align with the frozen pre-trained encoder's output and (2) fine-tuning the encoder for an inpainting task based on a frozen decoder's input. Evaluations are conducted under single- and multi-speaker conditions using in-domain datasets and out-of-domain datasets (including unseen speakers, diverse speaking styles, and noise). Both informed and blind inpainting scenarios are considered, where the position of the corrupted segment is either known or unknown. The proposed SSL-based methods are benchmarked against several baselines, including a text-informed method combining automatic speech recognition with zero-shot text-to-speech synthesis. Performance is assessed using objective metrics and perceptual evaluations. 
The results demonstrate that both approaches outperform baselines, successfully reconstructing speech segments up to 200 ms, and sometimes up to 400 ms. Notably, fine-tuning the SSL encoder achieves more accurate speech reconstruction in single-speaker settings, while a pre-trained encoder proves more effective for multi-speaker scenarios.
This demonstrates that an SSL pretext task can transfer to speech inpainting, enabling successful speech reconstruction with a pre-trained encoder.

\end{abstract}




\end{frontmatter}



\section{Introduction}
\label{section:introduction}

In speech processing, inpainting aims at restoring a speech signal that has been ``locally'' and severely corrupted. This means that a segment of the waveform is either highly degraded (e.g., by clicks or clipping \citep{zavivska2020survey}), or missing (e.g., due to packet loss during transmission \citep{730750, thirunavukkarasu2015survey}). This type of degradation severely affects and can even completely alter the signal quality and intelligibility for the corrupted segment and its direct neighbourhood, while the rest of the signal remains intact. This contrasts with speech enhancement in noise \citep{loizou2007speech}, where a more or less stationary additive noise is present throughout the entire signal duration.     
Speech inpainting has been explored in the past few decades using a variety of techniques (see \Cref{section:related} for a detailed review of the literature). Initially, classical signal processing was applied to short corrupted segments~\citep{gunduzhan2001linear,Lagrange2005LongIO}. This was followed by the use of 
deep neural networks (DNNs)-based encoder-decoder architectures, which directly map incomplete signals to the complete ones, through fully-supervised training~\citep{kegler20_interspeech,zhao23d_interspeech}. 
In particular, the encoder processes the signal surrounding the corrupted segment, while the decoder generates the signal within the corrupted segment. 
Lastly,
speech inpainting have been considered as one ``downstream'' task of a model pre-trained with Self-Supervised Learning (SSL), i.e., by fine-tuning the model to the inpainting task~\citep{xue2023contrast,zhang2024td}.

Interestingly, speech inpainting shares notable similarities with SSL 
that uses masking as a pretext task. Transformer-based SSL encoders such as HuBERT \citep{Hsu2021}, wav2vec \citep{schneider2019wav2vec}, wav2vec2.0 \citep{Baevski2020}, and WavLM \citep{Chen2022}, learn rich representations by masking portions of the input, and by training the model to predict the representations of the masked input segments using information from the unmasked context. 
While this process does not occur directly in the signal domain but in a non-linearly projected embedding space,
both speech inpainting and SSL share the objective of leveraging regularities across multiple time scales and linguistic levels to infer the content of either a severely corrupted, missing or masked segment within an input speech signal. 
Having drawn this parallel, to the best of our knowledge, the SSL masking pretext task has never been evaluated as such in the context of speech inpainting. \emph{We thus hypothesise that the mask prediction task used in SSL can be directly applied to speech inpainting, without the need for fine-tuning the SSL encoder.}
\newline

\begin{figure*}
  \centering
  \includegraphics[width=0.65\linewidth]{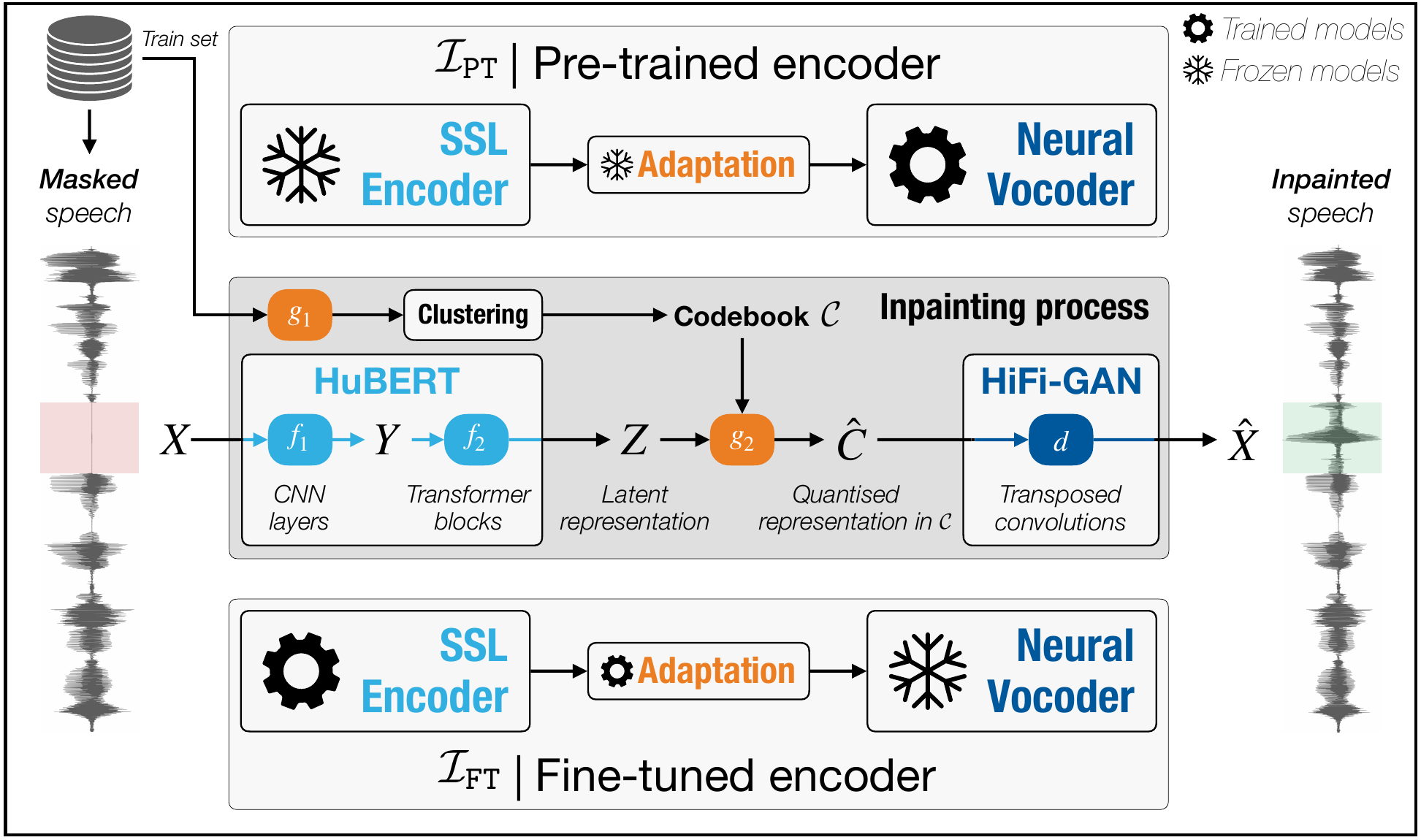}
  \caption{Proposed inpainting framework, leveraging a pre-trained SSL encoder on a unmasking task and using a neural vocoder for audio generation. Top: the SSL encoder is kept frozen (pre-trained) while the neural vocoder is adapted to the SSL output. Bottom: the SSL encoder is fine-tuned to an inpainting task, while the neural vocoder is kept frozen. In the present study we use HuBERT as the SSL and HiFi-GAN as the vocoder (subfigure in the middle). The inpainting process and adaptation mechanism between the SSL output and neural vocoder input are detailed in \Cref{section:method}).}
  \label{fig:framework}
\end{figure*}

To test this hypothesis, we investigate in this study the integration of an SSL encoder---specifically, HuBERT \citep{Hsu2021}---into a speech inpainting pipeline, utilising its rich contextual representations, spanning from phonetics to semantics \citep{10096149}, to “fill in the gap” by reconstructing the severely corrupted or missing portions of a speech signal based on its surrounding context.
More specifically, we utilise the HuBERT model for inference, maintaining the same configuration as in its SSL pre-training phase. We input signals that contain missing segments, and the model predicts the unmasked segments, which we then use to fill in the missing portions of the signals.
Since HuBERT does not directly predict the missing time-domain signal samples but rather high-dimensional embeddings, 
we thus combine the SSL encoder with a neural vocoder---in the present case HiFi-GAN \citep{kong2020hifi}. 
It is important to stress out that compared to \citep{xue2023contrast} and \citep{zhang2024td}, we do not fine-tune the SSL encoder. We separately train the neural vocoder to learn the mapping between the output of the pre-trained SSL model and the time-domain speech signal, as directly inspired from \citep{Polyak2021}. We denote this inpainting method \inpaintM (with $\M$ standing for pre-trained), which is summarised in the top of \cref{fig:framework}. 

For comparison, we then assess whether fine-tuning the SSL model to the inpainting task, i.e. treating inpainting as a downstream task instead of a pretext SSL training task, further enhances performance. We denote this inpainting method \inpaintI (with $\I$ standing for fine-tuned), as shown at the bottom of \cref{fig:framework}. In this case, we fine-tune the SSL encoder to fill in missing portions of the input signal directly in the audio domain. We choose a Mel-spectrogram representation for this purpose, as it can maintain the same temporal resolution as the original HuBERT representation space and serves as the standard input of a vanilla HiFi-GAN to convert back to the time-domain. Overall, let us note that our two approaches are symmetrical: \inpaintM involves a frozen pre-trained encoder with an adapted decoder, while \inpaintI involves a fine-tuned encoder with a frozen decoder.

Lastly, we compare the ability of an SSL encoder (specifically, HuBERT) for inpainting against a generative model. We introduce a baseline inspired by the text-informed speech inpainting study of \citet{7760374}. Their approach combines text-to-speech synthesis (TTS) and voice conversion (VC) to inpaint the corrupted portion of the audio signal, using the text from both uncorrupted and corrupted portions of the signal. We adapted their method to a textless inpainting scenario by incorporating an ASR frontend based on the Whisper model \citep{radford2023robust} to decode the most likely word sequence from the signal to be inpainted and replacing the TTS-followed-by-VC pipeline by a more recent zero-shot neural TTS. This baseline represents an additional contribution of this work. 
\newline

In this paper, we do not aim to propose a novel method for speech inpainting that outperforms state-of-the-art methods. Instead, we investigate whether the straightforward but so far unexplored solution of using a pre-trained SSL model as-is for speech inpainting is viable. To this end, we assess the performance of the proposed SSL-based methods in both single-speaker and multi-speaker settings, using in-domain datasets, which aligned with the training data, and out-of-domain datasets, which feature unseen speakers, varying speaking styles (including expressive speech), and different types of background noise---challenging conditions that, to the best of our knowledge, have received limited attention in previous studies. Performance is evaluated with both objective metrics and perceptual tests. 
To sum up, the contributions of this paper are:
\begin{enumerate}
    \item An extensive state-of-the-art of inpainting methods;
    \item The direct application of the mask prediction task used in SSL to speech inpainting (inpainting as a pretext task --- \inpaintM);
    \item The comparison with fine-tuning of SSL model to inpainting (inpainting as a downstream task --- \inpaintI);
    \item An extensive evaluation of both methods against different baselines and on in-domain and out-of-domain datasets;
    \item The complete source code\footnote{\url{https://gricad-gitlab.univ-grenoble-alpes.fr/huebert/speech-inpainting}} and a demo page for the two proposed speech inpainting frameworks and the different baselines.\footnote{\url{http://www.ultraspeech.com/demo/csl_2025_inpainting/}}
\end{enumerate}

\section{Related work}
\label{section:related}

\subsection{Speech inpainting}
\paragraph{Early signal processing approaches}

Speech inpainting studies initially focused on short corrupted segments (i.e., from a few milliseconds to a few tens milliseconds), using conventional digital signal processing approaches. The first targeted applications were packet loss recovery in telecommunications and streaming audio \citep{730750, thirunavukkarasu2015survey} or signal declicking \citep{zavivska2020survey}. 
Early works were based on signal processing techniques such as linear prediction / auto-regressive models \citep{502326, gunduzhan2001linear, kondo2006speech, zhang2008autoregressive}, Hidden Markov models \citep{rodbro2006hidden, borgstrom2010efficient}, sinusoidal modelling \citep{lindblom2002packet, Lagrange2005LongIO}, or graphs \citep{10.1109/TASLP.2018.2809864}.

\paragraph{Deep learning based approaches}

As already stated in the introduction, the speech inpainting problem has been tackled more recently with deep learning approaches. With the huge development of Voice on IP (VoIP) technologies, packet loss cancellation (PLC) has been a driving force application for deep learning-based speech inpainting studies. The domain has rapidly evolved in a few years considering longer corrupted segments (consecutive short-term frames), going from a few tens of milliseconds to several hundreds of milliseconds and more, and denser distributions of loss packets, with packet loss rates ranging from a few percent to more than \SI{50}{\percent}, possibly simulating a high level of VoIP degradation. 

In terms of architectures and training methodology, earlier studies have considered direct incomplete-to-complete signal mapping with a single network and fully-supervised training, most often using a few tens of hours of speech signal. This was applied in the time-frequency (TF) domain, using, e.g., a basic fully-connected neural network (FCNN) \citep{lee2015packet}, a U-Net architecture \citep{kegler20_interspeech, dai2024time}, a combination of U-Net with temporal convolutional network (TCN) bottleneck \citep{aironi2024complex}, or a convolutional recurrent network (CRN) \citep{zhang2024bs}. Other works have considered using a pair of 1D convolutional neural network (CNN) encoder and decoder for waveform analysis and reconstruction, e.g., \citep{shi2019speech, wang2021temporal, zhu2022packet, liu2022plcnet}. Finally, some studies have considered working directly on the time-domain waveform samples using a recurrent neural network (RNN), e.g., \citep{lotfidereshgi2018speech, mohamed2020concealnet, westhausen2022tplcnet}, or a CRN \citep{lin2021time}. Many works have considered the generative adversarial network (GAN) training approach \citep{goodfellow2014generative}, i.e., combining a generative network with a discriminative network at training time, for improving the quality of the inpainted speech signal, e.g., \citep{shi2019speech, wang2021temporal, pascual2021adversarial, liu2022plcnet, zhao23d_interspeech, aironi2024complex, dai2024time, zhang2024bs}. 

More complex encoder-decoder combinations have been proposed over the years. For example, a hybrid STFT-domain encoder combined with a time-domain CNN decoder was used in \citep{pascual2021adversarial}.  Taking inspiration from deep learning-based speech synthesis, and in particular the recent developments in text-to-speech synthesis using neural vocoders, speech inpainting and PLC studies have considered combining a missing features predictor network with a (possibly pre-trained) neural vocoder. For example, a GRU-based RNN designed to predict missing LPCNet input features was combined with an LPCNet vocoder in \citep{valin2022real}. A relatively basic FCNN-based Mel-spectrogram predictor taking as input the \num{11} previous frames (i.e., \SI{120}{\milli\second} context) was combined with a WaveGlow vocoder in \citep{zhou2022neural}. A full-band recurrent network (FRN) combining an encoder, a predictor and a joiner network was proposed in \citep{nguyen2023improving}. This FRN model was improved in \citep{yang2023towards} by adding a CTC-based ASR representation loss in the training, and in \citep{yang2023masked} by combining it with a masked frequency prediction for bandwidth extension. 
Recent works have integrated a Transformer module \citep{Vaswani2017} within the inpainting model, to exploit its natural ability to predict a (missing) sequence portion from a large temporal context. For example, the combination of CNN and Transformer was proposed in \citep{li2023time, zhao2024time} and a GAN TCN with a Conformer predictive network was proposed in \citep{zhao2023predictive}. \citet{li2022end} introduced a wav2vec-based perceptual loss term in the GAN training of their 1D CNN encoder-decoder model (in addition to the reconstruction loss and an ASR-based loss term). Very recently, a Flow Matching approach was proposed in \citep{yang2025flow}. 
The two PLC challenges~\citep{diener2022interspeech,diener2025icassp} summarise and compare the most recent inpainting methods for packet loss concealment. They have demonstrated excellent performance from the best systems in terms of intelligibility and naturalness of the resulting speech. 

Finally, closely related to our work, \citet{xue2023contrast} and \citet{zhang2024td} introduced an SSL representation at the encoder level and combined it with a neural decoder. \citet{xue2023contrast} implement an encoder with two parallel branches: one acoustic branch, made of 2D convolution and self-attention blocks, and one wav2vec2.0-like \citep{Baevski2020} semantic branch pre-trained with contrastive learning. The outputs of these two branches are merged and send to an HiFi-GAN vocoder~\citep{kong2020hifi}. Importantly, \citet{xue2023contrast} train the semantic branch from scratch using packet loss as the masking scheme, before training end-to-end the rest of the model with the same data. 
\citet{zhang2024td} propose a model similar to that of \citet{xue2023contrast}, though with an additional temporal-dilated module to fuse the wav2vec2.0-like semantic branch and the acoustic branch before decoding the output waveform with an in-house decoder. 
Overall, both studies use SSL models as ``feature extractors'' and perform end-to-end fine-tuning of their network on inpainting tasks. Conversely, we hypothesise that a pre-trained SSL encoder is inherently capable of performing an inpainting task, providing it with a decoder for domain adaptation.

\paragraph{Multimodal speech inpainting}

A few studies have investigated the use of a visual input such as the speaker's lips for guiding the speech inpainting process. This was implemented with an LSTM- or Transformer-based multimodal context encoder \citep{9413488, hsu2022revise}. Other speech inpainting studies have investigated text-informed methods, where the complete text (or phonetic information, or sequence of  linguistic targets) is provided, covering both the uncorrupted and corrupted parts of the signal, and used to drive the inpainting process \citep{7760374, borsos22_interspeech}. 
Text-informed speech restoration was also considered in \citep{koizumi2023miipher}. 
In this study, we do not consider any additional modality, we assume that only the locally-corrupted speech signal is available, and we thus focus on audio-only speech inpainting.

\subsection{Speech enhancement in noise, speech restoration, and speech coding}

As briefly stated in the introduction, speech enhancement in noise is a task that is related to speech inpainting in the sense that the speech signal is degraded and must be restored, but the nature of the degradation is significantly different. Therefore, the restoration process is also expected to be significantly different. However, after decades of development of speech enhancement methods working in the TF domain and using both conventional signal processing techniques \citep{loizou2007speech} and deep learning \citep{wang2018supervised}, 
recent studies in speech enhancement have proposed to use a combination of neural feature extractor with predictor taking the noisy speech as input and a neural vocoder for clean speech (re)generation, e.g., \citep{maiti2019parametric, du2020joint, liu2021voicefixer, su2021hifi, saeki2022selfremaster, irvin2023self, koizumi2023miipher}. This approach is closely similar to some of the above-mentioned inpainting studies, and brings to the speech enhancement problem the advantage of avoiding the delicate problem of signal phase reconstruction when only the magnitude spectrogram is denoised. 
We can note that HiFi-GAN is the preferred vocoder in \citep{saeki2022selfremaster, irvin2023self}. We can also note that in fact, the noisy feature extractor in \citep{irvin2023self, koizumi2023miipher} is an SSL model, making these studies closely related to our work in terms of model architecture and combination (HuBERT with HiFi-GAN is even among the multiple combinations tested by \citet{irvin2023self}). However, both studies  focus on speech enhancement in noise and dereverberation, and do not address inpainting specifically. Moreover, the model in \citep{koizumi2023miipher} also takes text as input, whereas our study focuses on audio-only inpainting.   

A few recent papers have reconsidered the speech enhancement problem as a generalized multi-task problem, possibly including denoising, dereverberation, bandwidth extension, declipping and packet loss concealment (hence inpainting) under the general denomination of speech restoration, see, e.g., \citep{liu2021voicefixer, saeki2022selfremaster, koizumi2023miipher}. However, in practice, in their experiments, these studies do not consider the inpainting task as defined in the present study, that is the recovery of long highly-corrupted or missing segments of speech. For example, \citet{saeki2022selfremaster} introduced their work in the context of speech restoration but only consider equalization and bandwidth extension in the experiments.

Finally, it can be noted that specifically combining HuBERT and HiFi-GAN was also proposed by \citet{Polyak2021} for speech coding, where the input signal is unaltered, and the primary objective is to generate an output signal that closely matches the input, while drastically limiting the bitrate of the encoded speech representation. As stated in the introduction, we inspired from this study but focus on speech inpainting, i.e., using the unmasking abilities of the SSL encoder in inference, which is a very different task than speech coding.

\subsection{Speech continuation with speech language models}

Speech continuation shares the task of predicting a missing portion of a speech signal with speech inpainting but is causal (prediction from a past context only), and operates over portion lengths that are orders of magnitude longer than those of conventional speech inpainting applications such as PLC. As a consequence, the most recent approaches leverage a generative speech language model (SpeechLM) for this task, i.e., a model  trained to generate speech via auto-regressive prediction and decoding of quantised elementary speech units referred to as speech ``tokens'' (see \citep{arora2025landscape} for a recent review). 
\citet{lakhotia2021generative} have demonstrated the ability of such models for audio-only speech continuation. \citet{borsos2023audiolm} and \citet{Chen2025} have further combined the prediction of audio tokens with ``semantic'' tokens that are derived from text-based language models. 
In addition to semantic tokens, \citet{Du2024} used ground-truth text as input for the missing speech portion to predict.

Overall, while those methods could be considered as suitable candidates for inpainting tasks, they do not fit the main hypothesis of this paper.
These methods leverage the generative ability of auto-regressive models for the speech continuation task while we take a step aside in studying the parallel between the unmasking pretext task of SSL encoders and speech inpainting. Moreover, the SSL-based pipeline investigated here is significantly lighter than recent SpeechLM approaches, which is essential for most practical use cases (e.g. embedded voice prothesis, real-time PLC).  
Nevertheless, as mentioned in introduction, we use a generative method as a baseline for comparison. We choose Whisper \citep{radford2023robust} which is based on a generative text-based language model, as text tokens are currently more efficient than audio tokens for next token prediction and therefore constitutes a stronger baseline than audio token-based methods.

\subsection{Conclusion}

In conclusion, SSL are becoming ubiquitous in a large variety of speech generation tasks, including speech coding, speech restoration, speech enhancement, speech continuation and speech inpainting.
Nevertheless, to the best of our knowledge, none of the related studies have considered an SSL encoder pre-trained on an unmasking task as a suitable candidate for speech inpainting without the need for fine-tuning. In the following Sections, we therefore present our implementation of a pre-trained HuBERT model in an inpainting pipeline, as well as its evaluation.

\section{Method}
\label{section:method}

\subsection{Problem formulation}

\begin{figure*}
  \centering
    \begin{subfigure}{0.533\linewidth}
    \includegraphics[width=\linewidth]{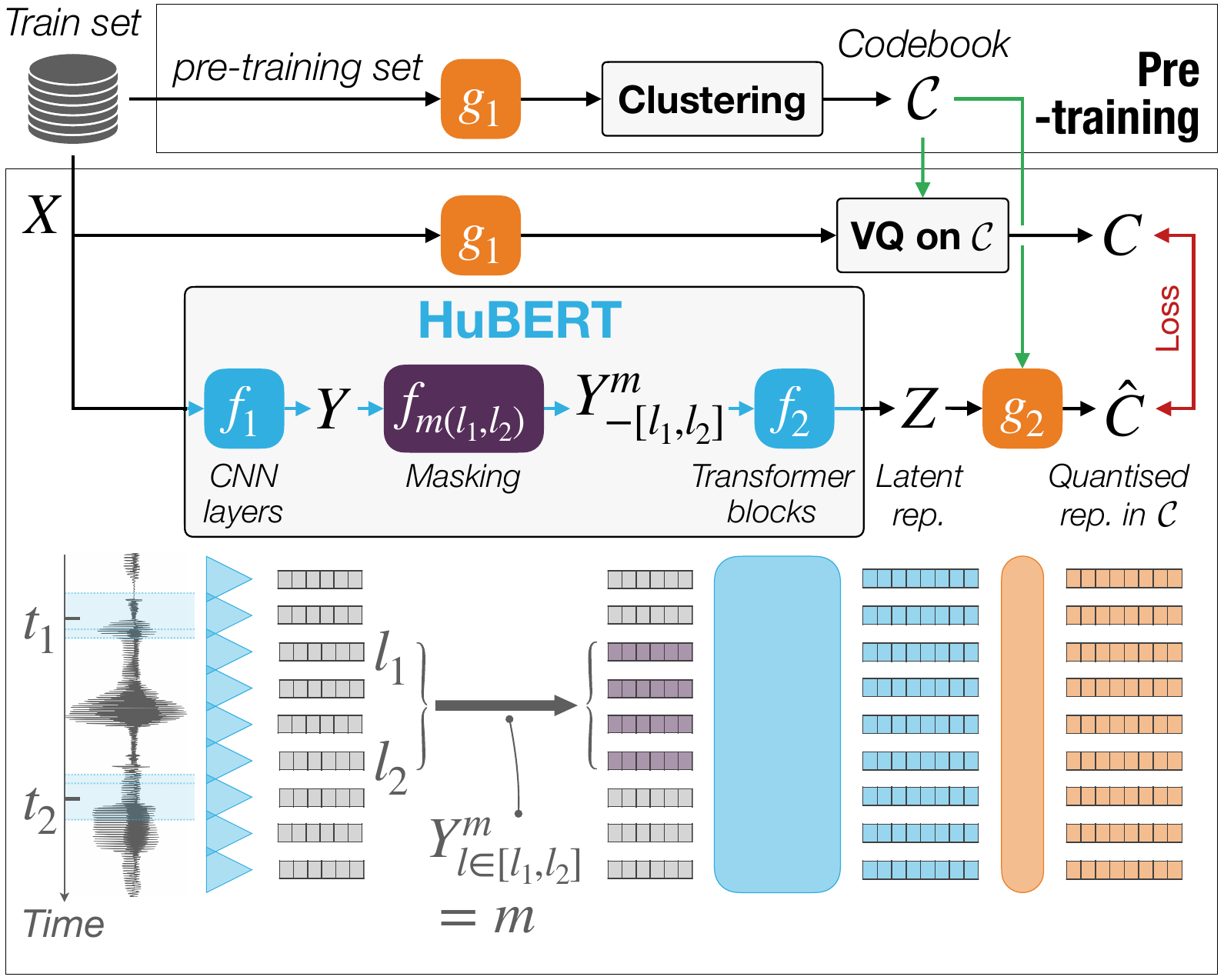}
        \caption{HuBERT training}
        \label{fig:hubert_train}
    \end{subfigure}
    \begin{subfigure}{0.312\linewidth}
    \includegraphics[width=\linewidth]{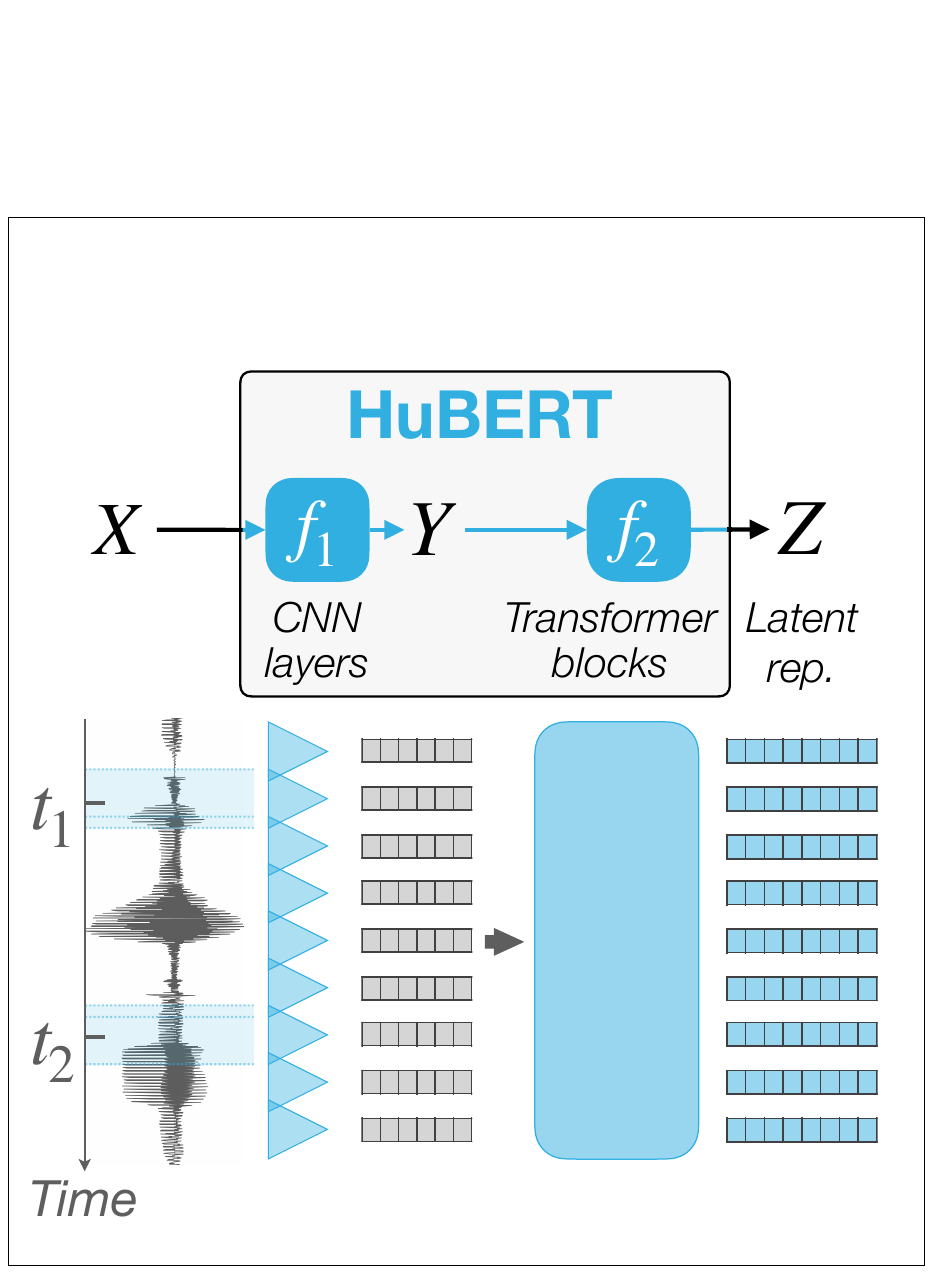}
        \caption{HuBERT inference}
        \label{fig:hubert_inf}
    \end{subfigure}
    \vspace{0.5cm}
    
    \begin{subfigure}{0.85\linewidth}
    \includegraphics[width=\linewidth]{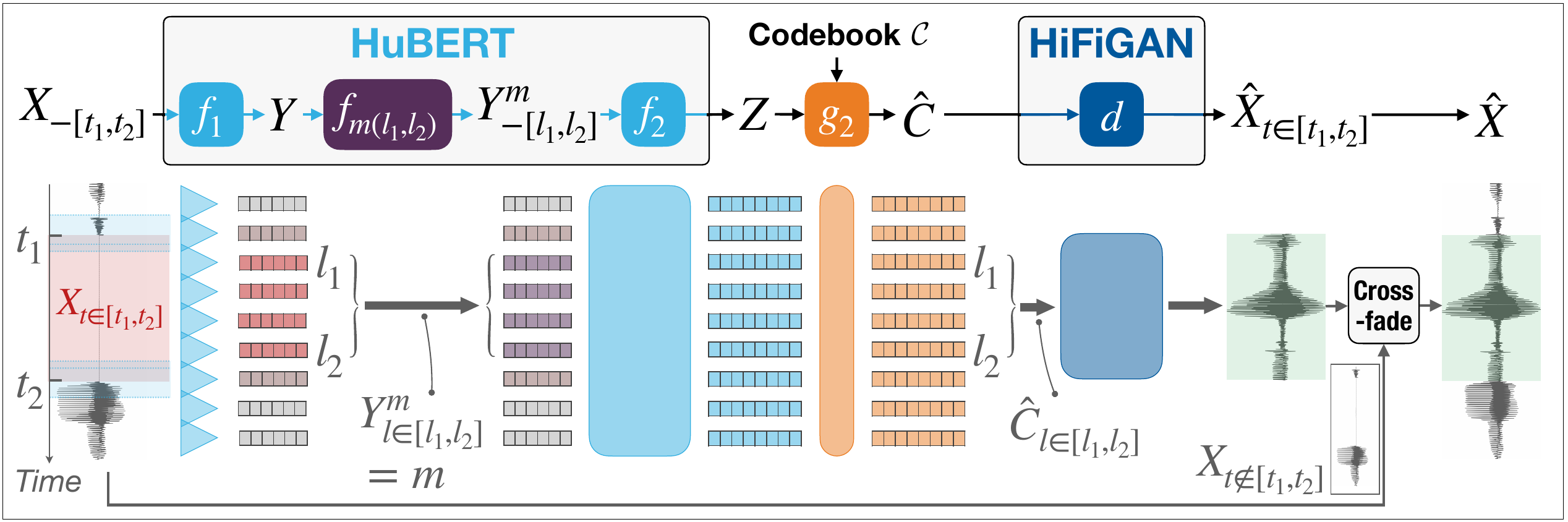}
        \caption{HuBERT-based inpainting pipeline -- informed case (inference)}
        \label{fig:informed}
    \end{subfigure}
    \vspace{0.5cm}
    
    \begin{subfigure}{0.85\linewidth}
    \includegraphics[width=\linewidth]{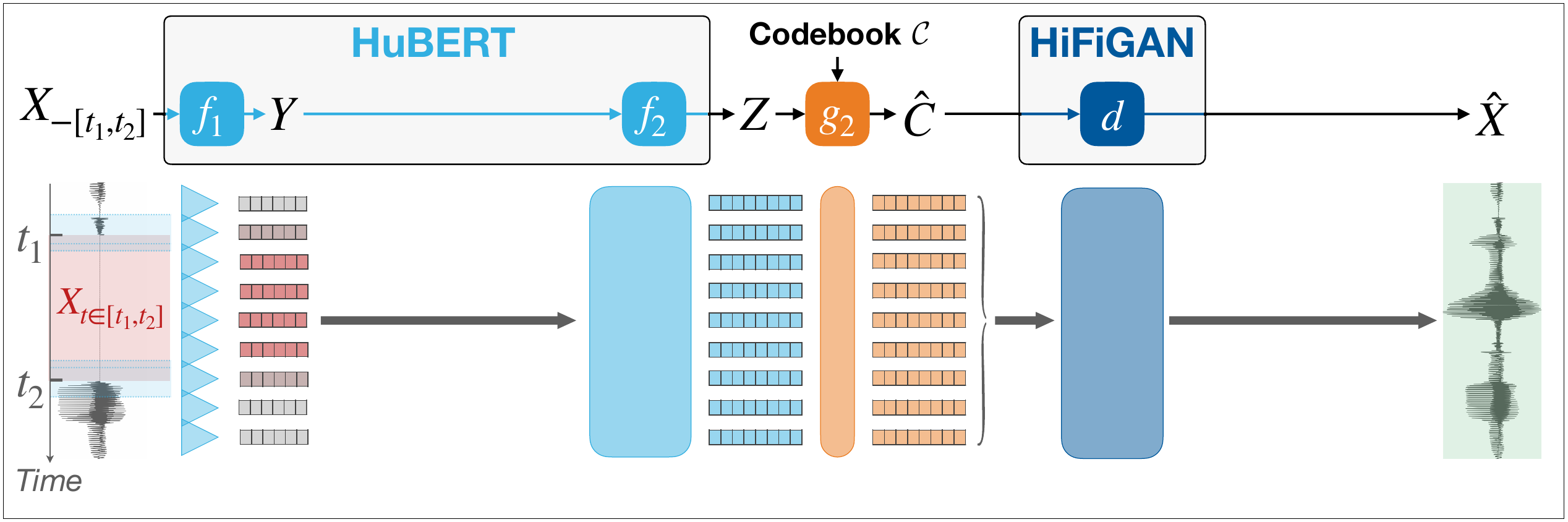}
        \caption{HuBERT-based inpainting pipeline -- blind case (inference)}
        \label{fig:blind}
    \end{subfigure}
    \vspace{0.5cm}

    \begin{subfigure}{0.85\linewidth}
    \includegraphics[width=\linewidth]{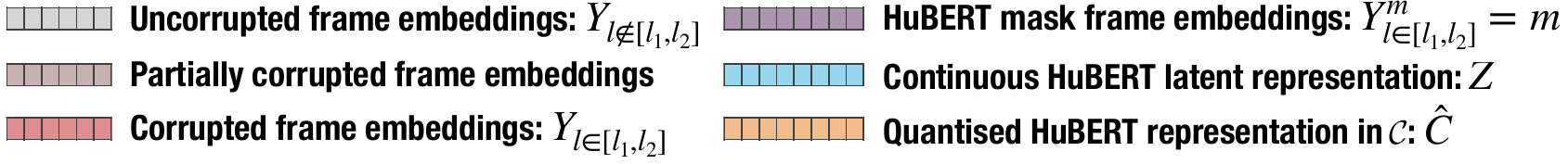}
    \end{subfigure}
    \vspace{0.5cm}
    
    \caption{Illustration of how the HuBERT model works during training (top-left) and inference (top-right). Adaptation of HuBERT for inference within the proposed inpainting pipeline, in the informed (middle) and blind (bottom) inpainting configurations.}
    \label{fig:informed-blind}
\end{figure*}

Following the notations introduced in \citep{Mohamed2022}, let $X = \left\{x_1,\ldots,x_T\right\}$ denote a sequence of speech samples of length $T$ (i.e., a waveform), and let $X_{-[t_1,t_2]}$ denote the sequence where the segment $X_{t\in[t_1,t_2]} = \left\{x_{t_1},x_{{t_1}+1},\ldots,x_{t_2}\right\}$ is corrupted (e.g., degraded or missing), and that we wish to reconstruct. 
In this paper, we focus on \emph{non-causal} inpainting, where the inpainting function \inpaint has access to both past and future uncorrupted portions of the input signal. We consider both the \emph{informed} inpainting paradigm, in which the corrupted segment position (i.e., the pair of values $\{t_1, t_2\}$) is known, and the \emph{blind} paradigm, in which the corrupted segment position is unknown. In the informed configuration, only the corrupted speech segment is predicted from its surrounding context. The uncorrupted portions of the signal are kept as is and are combined with the reconstructed segment using cross-fade at junctions to mitigate artifacts commonly associated with the concatenation of raw speech or audio segments. This process is illustrated in the right part of \cref{fig:informed}. It is formalised as:\footnote{For simplicity of presentation, we omit the cross-fade operation in this equation.}  
 \begin{align}
\hat{X}_{t\in[t_1,t_2]} = \mathcal{I}\left(X_{-[t_1,t_2]}\right) \ \ \text{and} \ \ \hat{X}_{t\notin[t_1,t_2]} = X_{t\notin[t_1,t_2]}.
\end{align}

In blind inpainting, the entire output signal is generated without distinguishing between the corrupted and uncorrupted parts, as illustrated in the right part of \cref{fig:blind}. This can be simply formalized as:
 \begin{align}
\hat{X} = \mathcal{I}\left(X_{-[t_1,t_2]}\right).  
\end{align}

\subsection{Masking, the core pretext task of HuBERT}
\label{subsec-foundation}

In this subsection, we briefly present the principles and core equations of HuBERT, as this formalism will be essential for explaining our inpainting pipeline in the subsequent subsections.
HuBERT is an encoder that transforms a speech signal $X$ of length $T$ into a latent continuous representation $Z = \left\{z_1,\ldots,z_L\right\}$, with dimension $D_z \times L$ \citep{Hsu2021}:
\begin{align}
y_l & = f_1\left(X_{[l H-u,l H+u[}\right), \forall l \in \{1,L\}, \label{eq:HuBERT_f1} \\
Z   & = f_2\left(Y\right), \ \ \text{with} \ \ Y = \left\{y_1,\ldots,y_L\right\}, \label{eq:HuBERT_f2}
\end{align}
where $f_1$, sometimes referred to as the prenet, is a stack of CNNs of span $2u$ samples and hop size $H$, and $f_2$ is a stack of Transformer encoder blocks. These core modules of HuBERT are illustrated in \cref{fig:hubert_inf} for inference. We do not detail their architecture in this paper, the reader is referred to the original HuBERT~\citep{Hsu2021} and Transformer papers~\citep{Vaswani2017} for details.
 
HuBERT is trained iteratively to predict a vector-quantised representation of the speech signal, denoted as $C$, from the latent representation $Z$. For instance, during the first  training iteration of the vanilla HuBERT, $C$ is a sequence of quantised Mel-frequency cepstral coefficient (MFCC) vectors.  
This is done with the help of two auxiliary modules, $g_1$ and $g_2$, which function as teacher and student models, respectively (see \cref{fig:hubert_train}):
\begin{itemize}
\item The teacher module $g_1$ maps the speech signal to the new representation (unquantised MFCC vectors in the above example), with a span of $2w$ samples and window shift $H$. Before training, a codebook $\mathcal{C}$ of quantised prototype vectors is created by passing a portion of the training set through $g_1$ and applying a k-means algorithm on the output. Then, during training, $g_1$ and vector quantisation (VQ) are used to extract a target quantised sequence $C = \left\{c_1,\ldots,c_L\right\}$ from each waveform of the train dataset:
\begin{eqnarray}
\label{eq:g1}
c_l & = & \texttt{VQ}_\mathcal{C} \big( g_1(X_{[lH-w,lH+w[}) \big), \forall l \in \{1,L\},
\end{eqnarray}
where $\texttt{VQ}_\mathcal{C}$ stands for VQ using the codebook $\mathcal{C}$.
\item The student module $g_2$ is designed to predict the quantised sequence $\hat{C}$ from the encoder output $Z$. In practice, $g_2$ is implemented using a softmax function, which involves a (learned) linear projection of $z_l$ onto embedding vectors representing the codebook prototypes.  
\end{itemize}
Importantly, during training, a portion of the CNN-encoded sequence $Y$ is randomly masked before being used to generate the fully-encoded sequence $Z$. Specifically, a sub-sequence  $Y_{l \in [l_1,l_2]}$ of $Y$, randomly selected, is replaced with ($l_2 - l_1 + 1$ occurrences of) a mask embedding vector $m$, which is jointly learned during the training process. This masking operation, spanning frames $l_1$ to $l_2$, is illustrated in \cref{fig:hubert_train}. We denote it as:
\begin{align}
Y^m_{-[l_1,l_2]} = f_{m(l_1,l_2)}(Y).
\end{align}
In summary, during HuBERT training, we have:
\begin{align}
y_l & = f_1\left(X_{[l H-u,l H+u[}\right), & \forall l \in \{1,L\}, \label{eq:HuBERT_f1-b} \\
Z   & = f_2 \circ f_{m(l_1,l_2)} \left(Y\right), & \ \ \text{with} \ \ Y = \left\{y_1,\ldots,y_L\right\} \ \ \text{and} \ \ Z = \left\{z_1,\ldots,z_L\right\},
\label{eq:HuBERT_f2-mask} \\
\hat{c}_l & = g_2\left(z_l\right), & \forall l \in \{1,L\}.
\label{eq:HuBERT_g2}
\end{align}
During HuBERT training, $g_1$ is fixed, while $f_1$, $f_2$ and $g_2$ are updated to minimise the distance between $\hat{C}$ and $C$, with part of the $Y$ sequence randomly masked across batches. Note that $g_1$ and $g_2$ are only used for HuBERT training and are generally discarded during inference when HuBERT is applied to a downstream task. 
Similarly, in conventional HuBERT usage, masking is only applied during training (see \cref{fig:hubert_train} and \eqref{eq:HuBERT_f2-mask}) and it is discarded during inference (see \cref{fig:hubert_inf} and \eqref{eq:HuBERT_f2}). Consequently, the pre-trained HuBERT consists solely of $f_2 \circ f_1 $, as defined in \eqref{eq:HuBERT_f1} and \eqref{eq:HuBERT_f2}, and a newly trained module dedicated to the downstream task is typically appended to $f_2$.

\subsection{Linking masking, prediction and inpainting}

The proposed HuBERT-based inpainting pipeline relies on the following general principle. As discussed in the previous subsection, HuBERT is trained to predict an efficient speech representation $Z$ from a $Y$ sequence containing one or several masked segments.
The pretext task used for training HuBERT shares similarities with inpainting in its underlying principles. Indeed, in both cases, the goal is to reconstruct an output sequence from an input sequence that includes missing, corrupted, or masked portions. The difference lies in the nature of the input and output: (i) During HuBERT training, the reconstructed/predicted output is the speech representation $Z$, whereas in inpainting, it is directly the signal $X$; and (ii) during HuBERT training, the missing input information corresponds to the masked part of the CNN-encoded sequence $Y$, whereas in inpainting, it is a segment of the input signal $X$. In summary, HuBERT is trained to perform a form of speech inpainting, but from the $Y$ space to the $Z$ space, and not directly within the input signal space. Consequently, we integrate HuBERT in a speech inpainting pipeline, addressing points (i) and (ii) as follows.
 
To address point (i), and drawing inspiration from recent advancements in low-bit-rate neural speech codecs \citep{Polyak2021}, we combine HuBERT as an encoder with the neural vocoder HiFi-GAN~\citep{kong2020hifi} as a decoder. HiFi-GAN, used here to reconstruct the inpainted speech waveform $\hat{X}$ from HuBERT’s output $Z$, demonstrated state-of-the-art performance in the latest text-to-speech synthesis benchmark~\citep{Perrotin2025}. Its necessary adaptation to the proposed inpainting pipeline is detailed in the next subsection.

For point (ii), we remark that, since the prenet temporal span is fixed, the corrupted input signal segment $X_{t\in[t_1,t_2]}$ we aim to inpaint corresponds to a corrupted subset of vectors $Y_{l\in[l_1,l_2]}$ in the CNN-encoded sequence $Y$. Therefore, inpainting the corrupted signal $X_{-[t_1,t_2]}$ relies on HuBERT's ability to predict $Z$ accurately from the  corresponding complete (corrupted) sequence $Y_{-[l_1,l_2]}$. However, leaving the corrupted subsequence $Y_{l\in[l_1,l_2]}$ as is in $Y$ is suboptimal because, during HuBERT training, the vectors composing this subsequence are replaced with the mask $m$ (see \Cref{subsec-foundation}). For HuBERT, this results in a mismatch between training (on masked segments) and inference (on corrupted segments, i.e., its use in the inpainting pipeline) conditions. To resolve this issue, we explicitly replace $Y_{l\in[l_1,l_2]}$ with the mask $m$ \textit{during inference}. In other words, we apply \eqref{eq:HuBERT_f2-mask} instead of \eqref{eq:HuBERT_f2} at inference/inpainting time. This approach is feasible, however, only in the informed case, where the time sample boundaries $(t_1, t_2)$ of the corrupted input speech segment are known, allowing us to deduce the corresponding frame boundaries $(l_1, l_2)$ (see the left part of \cref{fig:informed}). In the blind case, the location of $Y_{l\in[l_1,l_2]}$ is unknown and we cannot apply the mask (see the left part of \cref{fig:blind}). In this case, we thus stick to \eqref{eq:HuBERT_f2} at inference/inpainting time and have to accept that HuBERT functions under a train/inference mismatch.    
The blind inpainting configuration can be seen as a way to evaluate HuBERT's ability to infer an accurate speech representation from the surrounding context of the missing/corrupted speech information, even when this missing/corrupted speech information is not explicitly indicated by a mask embedding vector.   

Note that, as mentioned at the end of \Cref{subsec-foundation}, in the various uses of HuBERT reported in the literature, masking is applied only during training and is never used during inference. This is because, when using the speech representation $Z$ in downstream tasks, the input signal $X$ is generally \textit{not} corrupted. To the best of our knowledge, the speech inpainting pipeline considered in this paper represents the first instance where the masking process, which is central to HuBERT's training pretext task, is explicitly used \textit{at inference time} (in the informed configuration).

\subsection{Combining HuBERT encoder with HiFi-GAN decoder}

In this subsection, we present in detail how we combined the HuBERT encoder with the HiFi-GAN decoder. To this end, we first recall that during training on masked $Y$ sequences, HuBERT predicts $Z$ via a quantised representation  (see~\eqref{eq:HuBERT_g2}). Therefore, to carefully match the pretext training task during inpainting, we retain the auxiliary module $g_2$, which transforms $Z$ into the quantised sequence $\hat{C}$. $\hat{C}$ is then used as input to the HiFi-GAN decoder (denoted $d$) (see \cref{fig:framework,fig:informed-blind}).
In the informed case, we only use the subsequence $\hat{C}_{l\in[l_1,l_2]}$ (corresponding to the masked portion of $Y$) to reconstruct the missing speech segment. The latter is then combined with the uncorrupted parts of the input signal (see \cref{fig:informed}). This can be expressed as:\footnote{Again, for simplicity of presentation, we omit the cross-fade operation in this equation.} 
\begin{align}
\begin{dcases} 
    \hat{X}_{t\in[t_1,t_2]} & =  \ d(\hat{C}_{l\in[l_1,l_2]}) \\
    \hat{X}_{t\notin[t_1,t_2]} & = \ X_{t\notin[t_1,t_2]},
    \end{dcases}
 \end{align} 
 where again, $[l_1,l_2]$ is the frame interval corresponding to the corrupted segment of input signal sample interval  $[t_1,t_2]$. In the blind case, we simply have (see \cref{fig:blind}):
 \begin{equation}
    \hat{X} =  d(\hat{C}).
 \end{equation} 
In the following, we detail how to adapt $g_2$ to interface HuBERT with HiFi-GAN, and how to accordingly configure $g_1$ and the codebook $\mathcal{C}$ to train $g_2$. 
This adaptation depends on the two \methods presented in introduction: (i) using a pre-trained  
($\M$) HuBERT, where the HiFi-GAN decoder is trained to fit a frozen HuBERT encoder, and (ii) 
fine-tuning the SSL encoder
($\I$), where we fine-tune the HuBERT encoder to the inpainting task, with an output that fits the standard input of a frozen HiFi-GAN decoder. These methods are illustrated in \cref{fig:framework} and detailed below.

\subsubsection{Pre-trained SSL}
\label{sec:polyak}
In this first approach, we use a pre-trained HuBERT model\footnote{The term “pre-trained” here refers to both the SSL training with masking and the ASR fine-tuning, as discussed in Section \ref{sec:fine-tuned-ssl}.} and keep it frozen. To adapt the HiFi-GAN decoder to the frozen pre-trained HuBERT, we adopt the following two-step adaptation process. 
In the first step, we directly use $Z$ as the new signal representation (i.e., $g_1^{(\M)}$ is identical to the frozen pre-trained HuBERT $f_2 \circ f_1$). A new codebook $\mathcal{C}^{(\M)}$ is obtained by running the k-means algorithm  on the $Z$ sequences derived from a subset of the pre-training dataset. $g_2^{(\M)}$ is then simply the quantisation on $\mathcal{C}^{(\M)}$ of the encoded sequence $Z$ corresponding to any corrupted input sequence $X_{-[t_1,t_2]}$:
\begin{equation}
    \hat{C} = g_2^{(\M)}(Z) = \texttt{VQ}_{\mathcal{C}^{(\M)}}(Z). 
\end{equation}

In the second step, we trained a HiFi-GAN model $d^{(\M)}$ from scratch to reconstruct the speech waveform from HuBERT's output. 
Specifically, the decoder takes the index of each $\hat{c}_l$ in the codebook $\mathcal{C}^{(\M)}$ as input and learns a look-up table of embedding vectors, which are then fed into the standard HiFi-GAN architecture. 

By denoting with $^{\ast}$ the modules that are trained, the full inpainting pipeline \inpaintM can be summarized as follows. In the informed case, we have:
\begin{align}
   \hat{X}_{t\in[t_1,t_2]} & = \mathcal{I}_{\M}\left( X_{-[t_1,t_2]} \right) \nonumber \\ 
   &= d^{(\M)*} \circ \big\{ g_2^{(\M)} \circ f_2 \circ f_{m(l_1,l_2)} \circ f_1 ( X_{-[t_1,t_2]})\big\}_{l\in[l_1,l_2]},
\end{align}
and in the blind case, we have:
\begin{align}
   \hat{X} &= \mathcal{I}_{\M}\left( X_{-[t_1,t_2]} \right) \nonumber \\
   &= d^{(\M)*} \circ g_2^{(\M)} \circ f_2 \circ f_1 \left( X_{-[t_1,t_2]} \right).
\end{align}

\subsubsection{Fine-tuned SSL}
\label{sec:fine-tuned-ssl}
In this second approach, we use the standard (vanilla) HiFi-GAN decoder, which takes a Mel-spectrogram as input and remains frozen, while we adapt HuBERT. To fully understand our adaptation of HuBERT and the motivation behind it, we must revisit the conventional HuBERT training process, briefly outlined in \Cref{subsec-foundation}.

As described in details in \citep{Hsu2021}, after HuBERT is initially pre-trained on a masking pretext task using $g_1$ and $g_2$, it is subsequently fine-tuned on an ASR task.
In this stage, $g_1$ and $g_2$ are discarded, and $f_2$ is extended with a softmax layer for connectionist temporal classification of characters, and fine-tuned using a labelled speech dataset. 
In our $\I$ approach to inpainting, we instead aim at predicting signal frames to fill-in the gap, while aligning with HiFi-GAN's Mel-spectrogram input representation. To achieve this, we reintroduce $g_1$ and $g_2$ and perform a new round of training using the masking pretext task on a reasonable amount of data. In this iteration, the speech representation is replaced by the Mel-spectrogram. Altogether, this adaptation process can be seen an inpainting-oriented fine-tuning. 
Conceptually, it can be seen as partially “unlearning” ASR-specific abstractions in order to recover more detailed signal properties. Importantly, however, the model still benefits from HuBERT’s strong ability to encode the contextual structure surrounding the gap.

In a few more details, we define $g_1^{(\I)}$ as the extraction of Mel-spectrogram vectors from (frames of $2w$ samples of) the waveforms $X$. The codebook $\mathcal{C}^{(\I)}$ is obtained with the k-means algorithm applied on the output of $g_1^{(\I)}$ 
for a train set. Given an input sequence $X$, the teacher module $g_1^{(\I)}$ computes a Mel-spectrogram vector for each frame, which is assigned to its closest centroid in $\mathcal{C}^{(\I)}$ (as in~\eqref{eq:g1}). The student softmax-based $g_2^{(\I)}$ module of~\citep{Hsu2021} is also re-introduced, to predict a sequence $\hat{C}$ of quantised Mel-spectrogram vectors in $\mathcal{C}^{(\I)}$:
\begin{equation}
\label{eq:g2}
    \hat{c}_l = g_2^{(\I)}(z_l) = \operatorname*{argmax}_{c}\left(\frac{\exp\Big( \text{sim}\big( A z_l, e_{c}^{(\I)}\big) / \tau \Big)}{\sum_{c'=1}^{\text{Card}(\mathcal{C}^{(\I)})}\exp\left( \text{sim}\big( A z_l, e_{c'}^{(\I)} \big) / \tau \right)}\right),
\end{equation}
where $A$ is a linear projection, $\text{sim}(a, b)$ is the cosine similarity between $a$ and $b$, $e_c^{(\I)}$ is a learnt embedding of codeword $c\in\mathcal{C}^{(\I)}$, and $\tau$ is the logit scale factor~\citep{Hsu2021}, set to 0.1. 
Following HuBERT training described in \Cref{subsec-foundation}, $g_1^{(\I)}$ and $f_1$ are fixed, and $f_2$ and $g_2^{(\I)}$ are updated to minimise the cross-entropy loss between the predicted indices of Mel-spectrogram vectors $\hat{c}_l$ (softmax output of~\eqref{eq:g2}) and the indices of the quantised Mel-spectrogram vectors $c_l \in \mathcal{C}^{(\I)}$, while part of the $Y$ sequence is randomly masked across batches.
At inference, the sequence $\hat{C}$ of quantised Mel-spectrogram vectors is directly fed to a pre-trained vanilla HiFi-GAN. 

By noting again by $^{\ast}$ the modules that are trained, the full \inpaintI~framework can be summarized as follows.  In the informed case, we have:
\begin{align}
   \hat{X}_{t\in[t_1,t_2]} & = \mathcal{I}_{\I}\left( X_{-[t_1,t_2]} \right) \nonumber \\ 
   &= d \circ \big\{ g_2^{(\I)*} \circ f_2^{\ast} \circ f_{m(l_1,l_2)} \circ f_1 ( X_{-[t_1,t_2]} ) \big\}_{l\in[l_1,l_2]},
\end{align}
and in the blind case, we have:
\begin{align}
   \hat{X} &= \mathcal{I}_{\I}\left( X_{-[t_1,t_2]} \right) \nonumber \\
   &= d \circ g_2^{(\I)*} \circ f_2^{\ast} \circ f_1 \left( X_{-[t_1,t_2]} \right).
\end{align}

\section{Experimental set-up}
\label{section:experiments}

\subsection{Datasets}
\label{sec:datasets}
\paragraph{In-domain train/test sets} 
We conducted experiments on both LJ Speech~\citep{Ito2017} and VCTK~\citep{Yamagishi2019} datasets. LJ Speech is an English corpus containing \num{13100} short audio clips recorded by a single female speaker for a total length of approximately \SI{24}{\hour}. We isolated \num{12950} clips as the training/validation  set, the remaining \num{150} clips being used for test.   
VCTK includes a set of \num{43859} audio clips recorded by \num{109} English speakers  balanced in gender and with various accents, for a total of approximately \SI{44}{\hour}. We used \num{41747} clips from \num{105} speakers for training and \num{389} clips from \num{4} speakers for test. Importantly, we carefully designed the partitioning of the VCTK dataset to have no overlap between the training and test sets in terms of sentences and speakers. In other words, the proposed models have to generalise to both new speakers and new linguistic content. 

\paragraph{Out-of-domain test sets} We also considered two supplementary datasets for evaluating the extrapolation capabilities of our models to out-of-domain data. 
First, we designed noisy versions of our LJ Speech and VCTK test sets with the same number of utterances as described before (i.e., \num{150} for LJ Speech and \num{389} for VCTK). On each clip, we then applied either a white noise or a crowd noise with three signal-to-noise ratio (SNR) levels (\num{20}, \num{10} and \SI{0}{\decibel}) for a total of six noise-corrupted signals per utterance. Those two new test sets account for a total of \num{900} utterances for LJ Speech and \num{2334} for VCTK.

Second, we used a subset of the Expresso dataset~\citep{nguyen2023expresso}, a multi-speaker expressive speech dataset covering seven different speaking styles uttered by four different speakers. Following the recommendation from the authors of the Expresso dataset, our test set includes \num{588} utterances, with \num{147} clips per speaker and \num{84} clips per speaking style.

\subsection{Implementation details}
\label{sec:implementation}
\paragraph{SSL encoder HuBERT} 
For the two proposed inpainting frameworks \inpaintM~and \inpaintI, we used the HuBERT-large model \textit{hubert-large-ls960-ft}, publicly available on Hugging Face (see our repository). 
This model is a fine-tuned version of \textit{hubert-large-ll60k}, the latter being initially trained on the Libri-Light dataset \citep{librilight}, including \SI{60000}{\hour} of speech data from over \num{7000} speakers. The fine-tuning was done on the LibriSpeech dataset, containing \SI{960}{\hour} of speech data from over \num{2484} speakers. To the best of our knowledge, the LJ Speech and VCTK datasets used in the present study are not included in LibriSpeech. However, since LJ Speech is extracted from the LibriVox\footnote{\url{https://librivox.org}} 
dataset, there might be a slight overlap with the very large Libri-Light dataset (also based on LibriVox). However, this overlap is  approximately \num{0.0004}\% (\SI{24} vs.~\SI{60000} hours) and thus remains very limited. 

In the HuBERT model used in this study, the speech input is expected to be sampled at \SI{16}{\kilo\hertz}. The prenet window size and hop size are~\num{400} and \num{320} samples (i.e., \num{25} and \SI{20}{\milli\second}), respectively. The output $Z$ has dimension \num{768}.

\paragraph{Inpainting with pre-trained SSL (\textbf{\inpaintM})} 

For this approach, we used the implementation of the speech encoder-decoder framework proposed by~\citet{Polyak2021}. Dedicated codebooks $\mathcal{C}^{(\M)}$ were computed using the LJ Speech (resp. VCTK) dataset, considering a training subset of  \SI{21}{\hour} (resp. \SI{36}{\hour}). 
For the single-speaker setting, we choose 100 clusters, similarly to the first pre-training iteration of HuBERT (i.e., for k-means clustering of MFCC features)~\citep{Hsu2021}. For the multi-speaker setting, we found in preliminary experiments that increasing the codebook size to 500 did provide noticeable qualitative improvements, while offering a good trade-off between performance and computational cost. We therefore adopt this size. 
Recall that for this model $g_2^{(\M)}$ is not trained, i.e., for any representation $Z$ of a masked input sequence $X_{-[t_1,t_2]}$, $g_2^{(\M)}$ retrieves the closest sequence of vectors $\hat{C} = \left\{\hat{c}_1,\ldots,\hat{c}_L\right\}$ from $\mathcal{C}^{(\M)}$.
HiFi-GAN is then trained from scratch to generate $\hat{X}$ from $\hat{C}$. 
Note that contrary to~\citep{Polyak2021}, we did \emph{not} use a fundamental frequency ($f_o$) encoder as an input in parallel to the $\hat{C}$ sequence in order to avoid the explicit prediction of the $f_o$ of the corrupted segment. 
For the multi-speaker configuration (i.e., model trained on the VCTK dataset), a speaker embedding extracted using the speaker identification model proposed in~\citep{heigold2016end} was used as an additional conditioning vector to HiFi-GAN. Here, we trained this model on the same VCTK training subset than for the codebook computation (\SI{36}{\hour}).  Both the HiFi-GAN vocoder and the speaker identification model were trained with the Adam optimiser \citep{KingmaB14} over \num{200} epochs, with a batch size of \num{32} and a learning rate of \num{2e-4}.

\paragraph{Inpainting with fine-tuned SSL (\textbf{\inpaintI})}

In this second approach, the HuBERT encoder is fine-tuned to directly predict a quantised Mel-spectrogram representation, which is converted back to a time-domain audio signal with HiFi-GAN. More precisely, $g_2^{(\I)}$ is an additional linear layer designed to adapt the HuBERT output $Z$ to a quantised Mel-spectrogram domain, as defined in~\eqref{eq:g2}. The Transformer blocks
$f_2$ are fine-tuned, while $g_2^{(\I)}$ is trained from scratch. For this sake, the teacher $g_1^{(\I)}$ computes 80-dimensional Mel-spectrogram vectors from the input speech, with a window size of~\SI{46}{\milli\second} and a hop size of~\SI{20}{\milli\second}.
As for the \inpaintM~framework, dedicated codebooks $\mathcal{C}^{(\I)}$ were computed on the LJ Speech (resp. VCTK) training subset. We used \num{100} (resp. \num{500}) clusters, but with the k-means applied on Mel-spectrogram vectors obtained from $g_1^{(\I)}$.
For each training set (LJ speech or VCTK), fine-tuning $f_2$ and training $g_2^{(\I)}$ was done using the Adam optimiser over \num{100} epochs, with a batch size of \num{8} and a learning rate of \num{e-4}.

For the audio synthesis, we used a pre-trained HiFi-GAN model,\footnote{More specifically, we used the \texttt{UNIVERSAL\_V1} model (see our repository).} taking an 80-dimensional Mel-spectrogram as input, and generating a waveform at \SI{22.05}{\kilo\hertz}.\footnote{An upsampling of HuBERT's codebook, initially computed considering \SI{16}{\kilo\hertz} speech input, was therefore necessary.} 
While being optional, a slight fine-tuning of HiFi-GAN on quantised ground-truth Mel-spectrograms was found to be beneficial for the overall audio quality. This was done using Adam over \num{50} epochs, with a batch size of \num{8} and a learning rate of \num{e-4}.

\paragraph{Post-processing}

For the blind inpainting case, the reconstructed signal was generated entirely by the neural vocoder. For the informed case, we only kept the generated signal corresponding to the masked part and we placed it within the original (masked) signal using a cross-fade of \SI{5}{\milli\second} on both sides. Finally, the inpainted signals obtained with the  \inpaintI~framework were resampled to \SI{16}{\kilo\hertz} for a fair comparison with the other framework and baselines.

\subsection{Baselines}
\label{subsec-baselines}

\paragraph{Linear baseline} As a first baseline, we implemented a simple inpainting method based on linear interpolation (\inpaintLI). For a given masked signal, it consists in calculating its Mel-spectrogram (as done in \Cref{sec:implementation}) and replacing the masked frames with a linear interpolation between the last frame before the mask and the first frame after the mask. The interpolated Mel-spectrogram was then fed to the pre-trained HiFi-GAN vocoder to generate a \SI{22.05}{\kilo\hertz} waveform, which was then downsampled to \SI{16}{\kilo\hertz}.

\paragraph{Comparison with other studies} We also wanted to compare the proposed speech inpainting frameworks with other recently published methods such as \citep{kegler20_interspeech,zhao23d_interspeech,xue2023contrast} and \citep{zhang2024td}. Unfortunately, no source code is publicly available for these studies 
and we could not perform experiments with these methods with the exact same configurations as the ones used in the proposed frameworks. Nevertheless, since we use common metrics (detailed below), in \Cref{section:results} we compare our results with the ones reported in \citep{kegler20_interspeech,zhao23d_interspeech,xue2023contrast}, and \citep{zhang2024td} at least in terms of order of magnitude.

\begin{figure*}[t]
  \centering
  \includegraphics[width=0.75\linewidth]{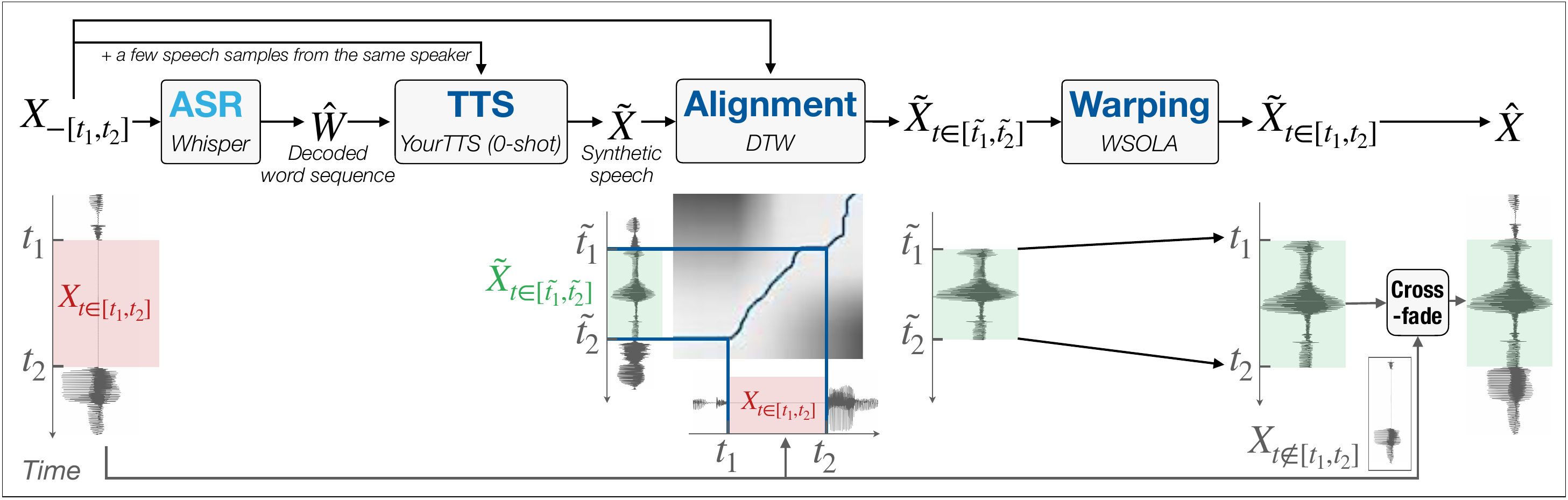}
  \caption{ASR-TTS baseline \inpaintT combining automatic speech recognition (ASR) and zero-shot text-to-speech (TTS) for text-informed speech inpainting. The corresponding inpainting process is detailed in \Cref{subsec-baselines}).}
  \label{fig:asr-tts}
\end{figure*}

\paragraph{ASR-TTS baseline}
In addition to the linear baseline and the results reported in comparable studies, we introduced an additional baseline based on generative modelling and inspired by text-informed speech inpainting, particularly the study by \citet{7760374}. 
For this study, we adapted their method to a text-less inpainting scenario by incorporating an automatic speech recognition (ASR) frontend to decode the most likely word sequence from the signal to be inpainted. This adaptation leverages the ASR’s internal language model to recover the linguistic content of the corrupted portion using its surrounding context. The resulting approach, denoted as \inpaintT, is presented in \cref{fig:asr-tts} and can be summarized as follows: 
\begin{enumerate}
    \item ASR is used to decode the most likely word sequence, $\hat{W}$, from the signal to be inpainted, $X_{-[t_1,t_2]}$.\footnote{We used the state-of-the-art \textit{whisper-large} model available on \textit{HuggingFace}; \url{https://huggingface.co/openai/whisper-large}}
    \item Zero-shot TTS  (also known as voice cloning) is used to synthesize $\tilde{X}$ using both $\hat{W}$ and the signal $X_{-[t_1,t_2]}$. The zero-shot TTS is preferred to the TTS-followed-by-VC approach  used in~\citep{7760374} since it allows to  generate a synthetic speech signal close to the target speaker's timbre  $\tilde{X}$ in a single processing step.\footnote{We used the \textit{YourTTS} system~\citep{pmlr-v162-casanova22a}, as implemented in the \textit{coqui-TTS} toolkit;  \url{https://github.com/coqui-ai/TTS}. Depending on the dataset used, if  $X_{-[t_1,t_2]}$ was too short to allow the TTS speaker encoder to effectively capture the speaker's timbre, additional utterances from the same speaker were concatenated to provide approximately 10 seconds of speech.} 
    \item The synthetic signal $\tilde{X}$ is time-aligned with the original signal $X_{-[t_1,t_2]}$ using Dynamic Time Warping (DTW). This alignment identifies which segment of the synthetic signal, $\tilde{X}_{t \in [\tilde{t_1}, \tilde{t_2}]}$, corresponds most closely to the missing chunk in the signal to inpaint, $X_{t \in [t_1, t_2]}$.
    \item The identified synthetic chunk is  stretched or compressed to match the target duration $(t_2 - t_1)$, replacing its original duration $(\tilde{t_2} - \tilde{t_1})$. This adjustment is performed using the Waveform Similarity-based Overlap-Add (WSOLA) algorithm~\citep{verhelst1993overlap}.
    \item The adjusted synthetic segment is inserted into $X_{-[t_1,t_2]}$ with cross-fade to ensure a seamless transition.
\end{enumerate}

Importantly, this approach is applicable only to the informed inpainting paradigm, as it requires knowledge of the exact boundaries of the portion to be inpainted ($t_1$ and $t_2$). It is also worth noting that this approach is relatively straightforward, as it does not involve any model training or fine-tuning; both the ASR and zero-shot TTS components are pre-trained and used “off-the-shelf.”

\subsection{Evaluation metrics}

\paragraph{Objective metrics}
We evaluated the inpainted speech quality using PESQ \citep{rix2001perceptual} and its intelligibility using STOI \citep{taal2010short} on both LJ Speech and VCTK test sets, comprising \num{150} and \num{389} utterances for the non-noisy versions and \num{900} and \num{2334} utterances for the noisy versions. To further evaluate the zero-shot learning capabilities of our models, we conducted the same experiments on the Expresso test set, consisting of \num{588} utterances. 
In all datasets, each utterance was masked three times using mask lengths of \num{100}, \num{200} and \SI{400} {\milli\second}, and whose position was randomly chosen in each utterance, resulting in a total of \num{4361} $\times$ \num{3} masked utterances to inpaint with our four frameworks (\inpaintLI, \inpaintT, \inpaintM, \inpaintI). 
PESQ and STOI were computed by considering segments of one second of speech, centred on the mask (therefore, the inpainted speech corresponds to \num{10}, \num{20}, or \SI{40}{\percent} of the original speech when measuring the score). 
As a complementary objective evaluation, we also performed automatic speech recognition (ASR) on the inpainted speech. We used the same pre-trained Whisper model~\citep{radford2023robust} as in our baseline \inpaintT and report the character error rate (CER).\footnote{This metric provides useful information about the phonetic content of the inpainted speech but may be biased by the linguistic prior on which the ASR may rely to transcribe it.} 
For all metrics, average scores on each test set and each mask length
are reported for all systems, with the binomial proportion confidence interval. 

\paragraph{Perceptual evaluation}
To further investigate the performance of the proposed inpainting system, we conducted an online MUSHRA-based listening test using the Web Audio Evaluation Tool~\citep{waet2015}. This was only done for the informed case. First, we randomly sampled 15 sentences from the LJ Speech test set (single-speaker condition), and 15 sentences from the VCTK test set (multi-speaker condition). For each sentence, we randomly masked a \SI{200}{\milli\second}-long segment and inpainted it with \inpaintI, \inpaintM, and with the \inpaintLI and \inpaintT baselines. Finally, we asked 100 native English speakers (self-reported as British or American, recruited via the Prolific platform\footnote{\scriptsize \url{https://www.prolific.co}}) to evaluate the quality of the inpainted speech using a MUSHRA-based protocol~\citep{series2014method}. For each sentence, presented in random order, participants had to rate comparatively the four inpainted signals as well as a high-anchor signal (natural uncorrupted speech). The type of each signal (natural or inpainted) was not given to participants.
As a reference, participants also received both the original uncorrupted sound file and its textual transcription. For each signal, they were instructed to focus on the inpainted speech segment which was highlighted using square brackets around the corresponding textual transcription, and asked to rate whether the highlighted speech segment was similar to that of the reference, on a continuous scale. Following the post-screening procedure described in \citep{series2014method}, we excluded 48 participants who ranked the hidden natural speech less than 90 out of 100, for more than 15\% of the stimuli (resulting in a final set of 52 participants who were considered as performing the test correctly).
Participants took a median time of \SI{19}{\minute} to complete the test and were compensated \num{3.75}~\pounds.

\paragraph{Statistical analysis}
In the following, we assess the effect of the \textit{mask length} (\num{100}, \num{200}, \SI{400} {\milli\second}), \textit{framework} (\inpaintLI, \inpaintT, \inpaintM, \inpaintI), \textit{dataset} (single- vs. multi-speaker) and \textit{type of inpainting} (informed vs. blind) factors, when relevant, on the objective and subjective metrics. The \textit{utterance} was added as a random factor. We each time use a beta regression model (using the R function \texttt{glmmTMB}), followed by post-hoc pairwise comparisons between factor levels (R function \texttt{glht}). Details of factors involved in each statistical analysis are given in the next Section. Significance level is systematically set to $p < 0.01$.

\section{Results}
\label{section:results}

\begin{figure*}[t!] 
  \centering
  \includegraphics[width=0.79\linewidth]{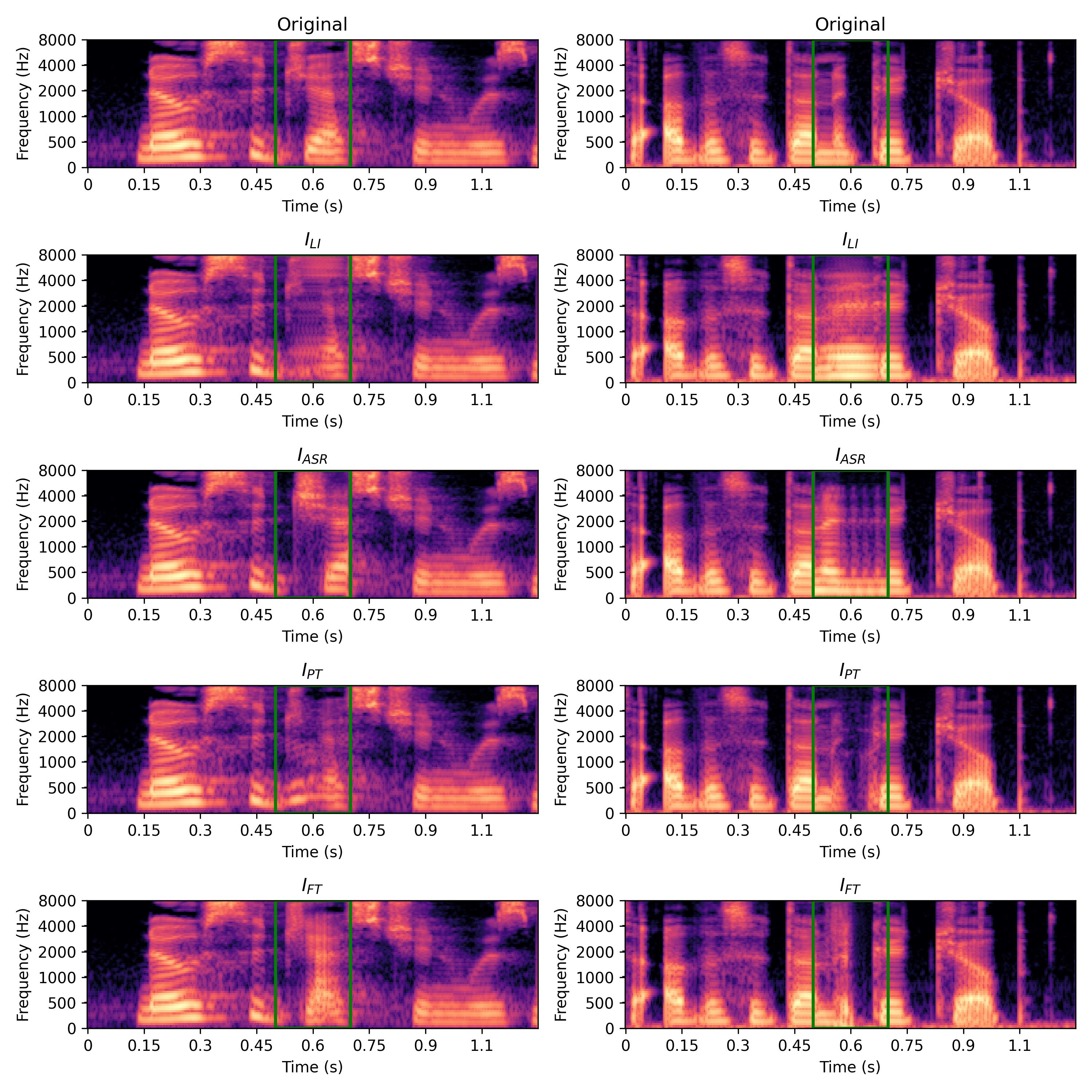}
  \caption{Examples of inpainted speech signals (80-dimensional Mel-spectrograms, informed case). Left: single-speaker for the sentence ``no su[gge]stion was made", Right: multi-speaker  for the sentence ``(...) has do[ne a goo]d job". The green rectangles illustrate the position and length of the mask (\SI{200}{\milli\second}).}
  \label{fig:ex_spectro}
\end{figure*}

\begin{table*}[t]
 \caption{Informed speech inpainting results. For all metrics, average scores with confidence intervals for each test set, each mask length, and each framework, for both the single- and multi-speaker configurations. The best scores per framework are denoted in bold. Pairs of symbols indicate pairs of distributions that are not significantly different.}
 \label{tab:inf}
 \centering
 \resizebox{\textwidth}{!}{
 \begin{tabular}{c|c|ccc||ccc}
 \multicolumn{2}{c}{} & \multicolumn{3}{c}{\textbf{Single-speaker (LJ Speech)}} & \multicolumn{3}{c}{\textbf{Multi-speaker (VCTK)}}\\
 \toprule
 Models & Mask (ms) & PESQ~$[-0.5;4.5]$ $\uparrow$ & STOI~$[0;1]$ $\uparrow$ & CER~(\%) $\downarrow$ & PESQ~$[-0.5;4.5]$ $\uparrow$ & STOI~$[0;1]$ $\uparrow$ & CER~(\%) $\downarrow$ \\ 
 \midrule 
 Unmasked   & 0 & $4.25 \pm 0.01$ & $0.97 \pm 0.00$ & $6 \pm 1$ 
        & $4.05 \pm 0.01$ & $0.94 \pm 0.01$ & $4 \pm 1$ \\
 \midrule 
        & 100 & $2.92 \pm 0.04$ \symbsigsquare{\colpurple} & $0.92 \pm 0.01$ & $13 \pm 2$ \symbsigcircle{\colpurple}
        & $2.40 \pm 0.02$ & $0.87 \pm 0.01$ & $17 \pm 1$ \\
Baseline \inpaintLI & 200 & $2.25 \pm 0.04$ & $0.83 \pm 0.01$ & $16 \pm 2$ \symbsigdiamond{\colpurple}
        & $1.97 \pm 0.02$ & $0.77 \pm 0.01$ & $13 \pm 1$ \symbsigtriangleu{\colpurple} \\
        & 400 & $1.95 \pm 0.03$ & $0.71 \pm 0.01$ & $19 \pm 1$ \symbsigstar{\colpurple}
        & $1.57 \pm 0.02$ & $0.60 \pm 0.01$ & $22 \pm 1$ \symbsigtriangled{\colpurple} \\ 
\midrule 
        & 100 & $2.97 \pm 0.05$ \symbsigsquare{\colpurple} & $0.93 \pm 0.01$ \symbsigcircle{\colred} \xspace \symbsigcircle{\colgreen} & $13 \pm 1$ \symbsigcircle{\colpurple}
        & $2.78 \pm 0.03$ & $0.90 \pm 0.01$ & $11 \pm 1$ \symbsigtriangleu{\colgreen}  \\
Baseline \inpaintT & 200 & $2.63 \pm 0.06$ & $0.88 \pm 0.02$  \symbsigcircle{\colred} \xspace \symbsigdiamond{\colgreen} & $17 \pm 1$ \symbsigdiamond{\colpurple}
        & $2.46 \pm 0.03$ & $0.85 \pm 0.01$ \symbsigtrianglel{\colgreen} & $13 \pm 1$ \symbsigtriangled{\colgreen} \xspace \symbsigtriangleu{\colpurple}  \\
        & 400 & $1.86 \pm 0.05$ & $0.82 \pm 0.02$ & $21 \pm 1$ \symbsigstar{\colpurple} \xspace \symbsigsquare{\colgreen}
        & $1.93 \pm 0.03$ & $0.79 \pm 0.01$ \symbsigtriangler{\colgreen} & $24 \pm 1$ \symbsigtriangled{\colpurple} \\ 
\midrule 
        & 100 & $3.06 \pm 0.04$ & $0.94 \pm 0.01$ \symbsigcircle{\colgreen}& $15 \pm 1$ \symbsigsquare{\colred}
        & $\mathbf{3.13 \pm 0.01}$ & $\mathbf{0.93 \pm 0.01}$ & $\mathbf{8 \pm 1}$  \\
\inpaintM  & 200 & $2.85 \pm 0.04$ & $0.89 \pm 0.01$ \symbsigtriangler{\collightblue} & $13 \pm 2$ \symbsigdiamond{\colorange} \xspace \symbsigsquare{\colred} \xspace \symbsigtriangled{\collightblue}
        & $\mathbf{2.93 \pm 0.01}$ & $\mathbf{0.88 \pm 0.01}$ \symbsigtriangler{\collightblue} & $15 \pm 1$  \symbsigtriangled{\collightblue}  \\
        & 400 & $2.78 \pm 0.04$ & $\mathbf{0.86 \pm 0.01}$ \symbsigstar{\colorange} & $24 \pm 3$ \symbsigsquare{\colgreen}
        & $\mathbf{2.66 \pm 0.02}$ & $\mathbf{0.83 \pm 0.01}$ & $\mathbf{18 \pm 1}$ \symbsigcircle{\colorange} \\
\midrule 
        & 100 & $\mathbf{3.28 \pm 0.01}$ & $\mathbf{0.96 \pm 0.01}$ & $\mathbf{7 \pm 1}$ 
        & $3.06 \pm 0.01$ & $0.90 \pm 0.01$ & $10 \pm 1$ \symbsigtriangleu{\colgreen}  \\
\inpaintI & 200 & $\mathbf{3.09 \pm 0.02}$ & $\mathbf{0.93 \pm 0.01}$ \symbsigdiamond{\colgreen}  & $\mathbf{12 \pm 1}$ \symbsigdiamond{\colorange}  \symbsigtriangleu{\collightblue}
        & $2.70 \pm 0.02$ & $0.85 \pm 0.01$ \symbsigtrianglel{\colgreen} & $\mathbf{12 \pm 1}$ \symbsigtriangled{\colgreen}\xspace \symbsigtriangleu{\collightblue} \\
        & 400 & $\mathbf{2.93 \pm 0.03}$ & $\mathbf{0.86 \pm 0.01}$ \symbsigstar{\colorange} & $\mathbf{14 \pm 1}$ 
        & $2.39 \pm 0.02$ & $0.79 \pm 0.02$ \symbsigtriangler{\colgreen} & $19 \pm 1$ \symbsigcircle{\colorange} \\
\midrule
\bottomrule
\end{tabular}
}
\end{table*}

\begin{small}
\begin{table*}[t]
 \caption{Blind speech inpainting results. For all metrics, average scores with confidence intervals for each test set, each mask length, and each framework, for both the single- and multi-speaker configurations. The best scores per framework are denoted in bold. Pairs of symbols indicate pairs of distributions that are not significantly different.}
 \label{tab:blind}
 \centering
 \resizebox{\textwidth}{!}{
 \begin{tabular}{c|c|ccc||ccc}
 \multicolumn{2}{c}{} & \multicolumn{3}{c}{\textbf{Single-speaker (LJ Speech)}} & \multicolumn{3}{c}{\textbf{Multi-speaker (VCTK)}}\\
 \toprule
 Models & Mask (ms) & PESQ~$[-0.5;4.5]$ $\uparrow$ & STOI ~$[0;1]$ $\uparrow$ & CER~(\%) $\downarrow$ & PESQ~$[-0.5;4.5]$ $\uparrow$ & STOI ~$[0;1]$ $\uparrow$ & CER~(\%) $\downarrow$ \\ 
 \midrule
\midrule 
        & 0 & $2.87 \pm 0.02$ & $0.89 \pm 0.01$ & $19 \pm 2$ 
        & $3.11 \pm 0.01$ & $0.93 \pm 0.01$ & $13 \pm 1$ \\
        & 100 & $2.77 \pm 0.04$ & $0.88 \pm 0.01$  \symbsigtriangled{\collightblue} & $40 \pm 3$ 
        & $\mathbf{2.93 \pm 0.01}$ & $\mathbf{0.89 \pm 0.01}$ \symbsigtriangled{\collightblue} & $26 \pm 1$  \\
\inpaintM & 200 & $2.33 \pm 0.04$ \symbsigtrianglel{\collightblue} & $0.75 \pm 0.01$ & $57 \pm 3$ 
        & $\mathbf{2.31 \pm 0.02}$ \symbsigtrianglel{\collightblue} & $\mathbf{0.71 \pm 0.01}$ & $\mathbf{31 \pm 1}$  \\
        & 400 & $1.72 \pm 0.03$ & $0.54 \pm 0.01$ \symbsigtriangler{\collightblue} & $81 \pm 4$ 
        & $\mathbf{1.53 \pm 0.02}$ & $\mathbf{0.52 \pm 0.01}$ \symbsigdiamond{\colorange} \xspace \symbsigtriangler{\collightblue} & $\mathbf{51 \pm 1}$ \\
\midrule 
        & 0 & $3.46 \pm 0.01$ & $0.95 \pm 0.01$ & $15 \pm 1$ 
        & $2.78 \pm 0.01$ & $0.89 \pm 0.01$ & $16 \pm 1$ \\
        & 100 & $\mathbf{2.81 \pm 0.03}$ & $\mathbf{0.90 \pm 0.01}$ & $\mathbf{17 \pm 2}$ 
        & $2.57 \pm 0.02$ & $0.81 \pm 0.01$ & $\mathbf{20 \pm 1}$  \\
\inpaintI & 200 & $\mathbf{2.55 \pm 0.04}$ & $\mathbf{0.84 \pm 0.01}$ & $\mathbf{24 \pm 3}$ 
        & $2.23 \pm 0.02$ & $0.69 \pm 0.01$ & $41 \pm 3$ \\
        & 400 & $\mathbf{1.97 \pm 0.03}$ & $\mathbf{0.79 \pm 0.01}$ & $\mathbf{39 \pm 2}$ 
        & $1.39 \pm 0.03$ & $0.51 \pm 0.01$ \symbsigdiamond{\colorange} & $56 \pm 3$ \\
   \bottomrule
 \end{tabular}
}
\end{table*}
\end{small}

\subsection{Qualitative results}

Examples of inpainted speech signals obtained with the two proposed frameworks (\inpaintI and \inpaintM) and with the baselines \inpaintLI and \inpaintT, in the informed case, and for a mask length of \SI{200}{\milli\second}, are presented in \cref{fig:ex_spectro}. Other examples, for other mask lengths and datasets, are available on our demo webpage\footnote{   \url{http://www.ultraspeech.com/demo/csl_2025_inpainting/}}.
We first examine the spectral pattern observed for the linear baseline \inpaintLI. Recall that the Mel-spectrogram is computed from the audio output of the HiFi-GAN vocoder, the latter being fed with a linearly interpolated mel-spectrogram between the beginning and the end of the mask. Interestingly, despite this ``linear input'', the inpainted speech is almost ---but not entirely--- stationary. In fact, for the single-speaker case (left column), we can observe a transient a few milliseconds after the start of the mask. Consequently, the neural vocoder has ``shaped'' the  linear input (likely not seen in its training corpus), probably by exploiting contextual information. However, as our quantitative evaluation confirms (see \Cref{sec:informed_objective}), this minimal sound shaping is not precise enough to recover the phonetic content of the masked part and the speech inpainted by the \inpaintLI framework is most often not intelligible.  

We now qualitatively compare the two SSL-based frameworks \inpaintI and \inpaintM and the \inpaintT baseline. For the single-speaker case (left column), the signal to be reconstructed corresponds to approximately two phones: a  post-alveolar affricate  /\textdyoghlig/ followed by a vowel /\textepsilon/ (in the word \textit{suggestion}). The complex spectral pattern associated with this phonetic sequence is better reconstructed by  \inpaintI and  \inpaintT frameworks than with \inpaintM, with a sharper vowel-consonant transition (\inpaintM wrongly maintains a strong formants structure during the consonant).  
For the multi-speaker case (right column), the signal to inpaint corresponds to the phonetic sequence [\textipa{n} \textipa{@} \textipa{g} \textipa{U}]. Here, \inpaintI is less efficient. It correctly reconstructs the initial nasal \textipa{n} as well as the plosive \textipa{g} and the final vowel \textipa{U} but surprisingly replaces the middle schwa \textipa{@} with an unvoiced and high energy sound, creating a kind of audio artefact. This is not the case with the \inpaintM framework, with which the signal is very well reconstructed. Regarding \inpaintT, the phonetic content is well reconstructed, as the linguistic targets are accurately decoded by the ASR. However, the transition between \textipa{@} and \textipa{g} sounds less natural than with \inpaintM, likely due to the absence of a clear closure before \textipa{g}, as indicated by the sound energy just before the burst of \textipa{g}. These initial qualitative results are confirmed by the quantitative evaluation presented in the following sections.

\subsection{Informed inpainting}

\subsubsection{Objective evaluation}
\label{sec:informed_objective}

The results of the objective evaluation of informed inpainting in terms of PESQ, STOI and CER scores are presented in \Cref{tab:inf}. We assessed here the significance of \textit{mask length} (\SI{100}, \SI{200}, \SI{400}{\milli\second}), \textit{framework} (\inpaintLI, \inpaintT, \inpaintM, \inpaintI) and \textit{dataset} (single- and multi-speaker) with the test utterances as a random factor for each objective metric. Statistical analysis showed that all factors and all their interactions have a significant effect on each objective metric. Non-significant pairs of distributions shown by post-hoc analyses are indicated by pairs of symbols in \Cref{tab:inf} and reported accordingly in the text.
\newline

\paragraph{Influence of mask length}
Pairwise comparisons show significant differences between the three \textit{mask length} levels, on all metrics, for each \textit{framework} and each \textit{dataset}, except in terms of STOI between mask lengths of \SI{100}{\milli\second} and \SI{200}{\milli\second} in the \inpaintT $\times$ single-speaker condition (\symbsigcircle{\colred}\normalsize); and in terms of CER between mask lengths of \SI{100}{\milli\second} and \SI{200}{\milli\second} in the \inpaintM $\times$ single-speaker condition (\symbsigsquare{\colred}\normalsize).
As expected, as the mask length increases from \SI{100}~ to \SI{400}{\milli\second}, the performance across all evaluated metrics decreases. For example, for the \inpaintI~framework, the PESQ score is \num{3.28} for a mask length of \SI{100}{\milli\second} and it drops to \num{2.93} for a mask length of \SI{400}{\milli\second}. 

\paragraph{Comparison between models and baselines} Pairwise differences between the four inpainting \textit{frameworks} metric distributions are as follows.
First, \inpaintM and \inpaintI display significant differences for all \textit{mask length} and \textit{datasets}, except in terms of STOI in the \SI{400}{\milli\second} $\times$ single-speaker (\symbsigstar{\colorange}\normalsize) condition; and in terms of CER in the \SI{200}{\milli\second} $\times$ single-speaker (\symbsigdiamond{\colorange}\normalsize) and in the \SI{400}{\milli\second} $\times$ multi-speaker (\symbsigcircle{\colorange}\normalsize) conditions. The nature of these differences depends on the \textit{dataset} and are discussed in the next Section.

Second, the baselines \inpaintLI and \inpaintT display significant differences for all \textit{mask lengths} and \textit{datasets} in terms of PESQ, except in the \SI{100}{\milli\second} $\times$ single-speaker (\symbsigsquare{\colpurple}\normalsize) condition; and for all \textit{mask length} and \textit{datasets} in terms of STOI. By contrast, they only differ in terms of CER in the \SI{100}{\milli\second} $\times$ multi-speaker condition.
This shows that if \inpaintT is superior in terms of quality of the signal reconstruction, both baseline methods are similar in terms of intelligibility of the reconstructed signal.

Third, both proposed inpainting frameworks (\inpaintI~and \inpaintM) obtain scores that are systematically greater (and often much greater) than those obtained by the \inpaintLI baseline, for both the single-speaker and multi-speaker cases.
This confirms the expected need for a non-linear modelling to fill gaps that include more than a diphone transition. 
This also demonstrates the interest of using a powerful encoder like HuBERT, which is able to exploit the contextual information  to access the high-level linguistic information needed for inpainting long gaps. 

Finally, both proposed inpainting frameworks (\inpaintI~and \inpaintM) obtained significantly higher scores than \inpaintT in terms of PESQ for all conditions. Both methods also outperform the \inpaintT baseline in terms of STOI and CER in the single-speaker setting, expect between the \inpaintT and \inpaintM STOI in the \SI{100}{\milli\second} $\times$ single-speaker condition (\symbsigcircle{\colgreen}\normalsize); between the \inpaintT and \inpaintI STOI in the \SI{200}{\milli\second} $\times$ single-speaker condition (\symbsigdiamond{\colgreen}\normalsize), and between the \inpaintT and \inpaintM CER in the \SI{400}{\milli\second} $\times$ single-speaker condition (\symbsigsquare{\colgreen}\normalsize).
In the multi-speaker setting, only \inpaintM is competitive compared to \inpaintT, displaying significantly greater score for both STOI and CER in all conditions.
These results highlight the effectiveness of the SSL-based approach, particularly \inpaintM. It can be hypothesized that HuBERT successfully infers the linguistic content of missing signal segments from their surrounding context, achieving performance comparable to the ASR used in \inpaintT, especially in the \SI{100}{\milli\second} and \SI{200}{\milli\second} conditions, where the inpainted segments are shorter than a word. However, the performance of both approaches declines for masks of approximately \SI{400}{\milli\second}, which roughly corresponds to the duration of a word. In such cases, the target word is often omitted in the ASR output, leading to poor signal reconstruction. Additionally, the SSL-based approaches appear to generate fewer audio artifacts compared to \inpaintT, even when the target text is accurately decoded. This may be due to challenges faced by the zero-shot TTS in accurately replicating the target speaker’s timbre, as well as potential alignment errors during the DTW.

\paragraph{Single-speaker vs.~multi-speaker} 
All metrics distributions between \textit{datasets} are also significant for each \textit{framework} and \textit{mask length}, except between the single- and multi-speaker datasets in terms of STOI in the \inpaintM $\times$ \SI{200}{\milli\second} (\symbsigtriangler{\collightblue}\normalsize) condition; and in terms of CER in the \inpaintM $\times$ \SI{200}{\milli\second} (\symbsigtriangled{\collightblue}\normalsize) and in the \inpaintI $\times$ \SI{200}{\milli\second} (\symbsigtriangleu{\collightblue}\normalsize) conditions.
Interestingly, results display a strong interaction between the \textit{dataset} and the \textit{framework} factors.
In the single-speaker case, 
the \inpaintI~framework (fine-tuned SSL encoder) consistently outperforms the \inpaintM~framework (frozen pre-trained SSL encoder) across all evaluated metrics. For example, for a mask length of \SI{100}{\milli\second}, \inpaintI~achieves a PESQ score of \num{3.28}, a STOI score  of \num{0.96}, and a CER of \SI{7}{\percent}, whereas \inpaintM~obtains \num{3.06}, \num{0.94}, and \SI{15}{\percent}, respectively. Conversely, in the multi-speaker setting (VCTK dataset), best performances are systematically obtained with \inpaintM. For example, 
with a mask length of \SI{400}{\milli\second}, \inpaintI~gets a PESQ score of \num{2.39}, a STOI score of \num{0.79}, and a CER of \SI{19}{\percent}, whereas \inpaintM~yields scores of \num{2.66}, \num{0.83}, and \SI{18}{\percent} respectively.
The difference probably stems from the difficulty for \inpaintI~to compress in a single codebook all the inter-speaker variability. The use of a speaker embedding as done in \inpaintM~appears to be a much more efficient strategy.

\subsubsection{Perceptual evaluation}

\begin{figure}[t]
  \centering
  \includegraphics[width=0.7\linewidth]{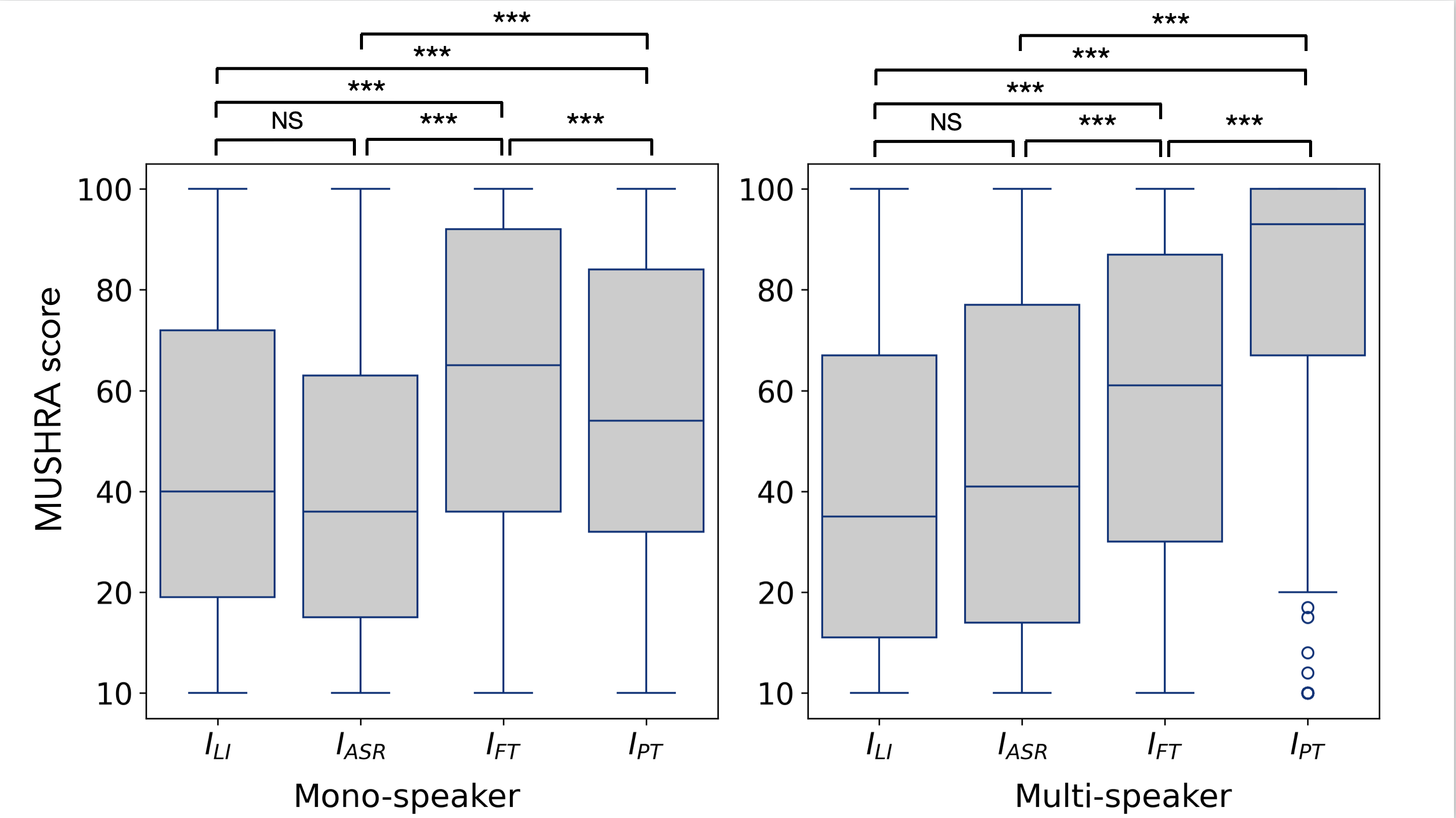}
  \caption{Boxplots illustrating the distribution of the MUSHRA scores for the two models \inpaintI~and \inpaintM, and for the \inpaintLI and \inpaintT baselines (informed inpainting with a \SI{200}{\milli\second}-length mask). *** indicates that the differences between each pairs of inpainting frameworks were found very significant (i.e. $p \leq 0.001$).}
  \label{fig:MUSHRA}
\end{figure}

Results of the perceptual evaluation, conducted in the informed case with masks of \SI{200}{\milli\second}, are presented in \cref{fig:MUSHRA}. We assessed here the significance of the \textit{framework} (\inpaintLI, \inpaintT, \inpaintM, \inpaintI) factor with the participants as a random effect, and pairwise comparison displays significant differences between all frameworks.
These results confirm all the trends revealed by the objective scores. First, despite its superior performance in terms of sound quality (see PESQ scores), the \inpaintT baseline is not significantly preferred by listeners over the \inpaintLI approach. Second, the two SSL-based frameworks \inpaintI and \inpaintM clearly outperform these baselines. The \inpaintI~framework provides better results than the \inpaintM~framework in the single-speaker case, and the opposite is observed in the multi-speaker case (with an even more marked difference between the two frameworks). Notably, the \inpaintM framework achieves a high performance level in the multi-speaker case, exceeding 90\%, with the reconstructed signal closely resembling the original.
\newline

\subsection{Blind inpainting}
The results of the objective evaluation of blind inpainting in terms of PESQ, STOI and CER scores are presented in \Cref{tab:blind}. 
To compare informed vs. blind inpainting, we assessed the significance of \textit{mask length} (\num{100}, \num{200}, \SI{400}{\milli\second}), \textit{framework} (\inpaintM, \inpaintI), \textit{dataset} (single- and multi-speaker) and \textit{type of inpainting} (informed vs. blind) with the test utterances as a random factor for each objective metric. Note that compared to \Cref{sec:informed_objective}, the \inpaintLI and \inpaintT levels are removed from the \textit{framework} factor, as they are not evaluated in the blind inpainting case.
Statistical analysis shows that all factors and all their interactions have a significant effect on each objective metric.
\newline

\paragraph{Informed vs.~blind} 
Pairwise comparisons show significant differences between the informed and blind metrics distributions, for each \textit{framework}, \textit{dataset}, and \textit{mask length}.
Compared to the informed inpainting configuration, the blind configuration is more challenging (the position of the mask is unknown and the full signal hence is reconstructed). As expected, it leads to lower performance, and this is observed for both \inpaintI~and \inpaintM, for both datasets, all mask lengths, and all metrics. For example, for blind inpainting with a 
\SI{200}{\milli\second} mask length in the single-speaker case, \inpaintI~gets a STOI score of \num{0.84}, compared to \num{0.93} in the corresponding informed case. Moreover, informed inpainting methods consistently exhibit lower CER, reflecting higher accuracy in reconstructing corrupted segments. 
\newline

\paragraph{Effect of mask length, framework, and dataset}
All pairs of distributions across the three factors are significant, except in terms of PESQ between single- and multi-speaker datasets in the \inpaintM $\times$ \SI{200}{\milli\second} condition (\symbsigtrianglel{\collightblue}\normalsize); in terms of STOI between the \inpaintM and \inpaintI frameworks in the \SI{400}{\milli\second} $\times $ multi-speaker condition (\symbsigdiamond{\colorange}\normalsize)  and between single- and multi-speaker in the \inpaintM $\times$ \SI{100}{\milli\second} (\symbsigtriangled{\collightblue}\normalsize) and \inpaintM $\times$ \SI{400}{\milli\second} (\symbsigtriangler{\collightblue}\normalsize) conditions.
Interestingly, the interactions between the \textit{type of inpainting} and each of these three factors are weak, as all the trends observed in the informed inpainting case remain in the blind inpainting case. Similarly to informed inpainting, performances of all metrics drop as mask length increases. Above all, the interaction between \textit{framework} and \textit{dataset} is still present, as the \inpaintI framework provides better results than the \inpaintM framework in the single-speaker case, and the opposite is observed in the multi-speaker case.

\subsection{Comparison with other studies} 
\label{sec:otherstudies}

As announced in \Cref{subsec-baselines}, we compare the overall performance of the proposed frameworks with that of recently published methods \citep{kegler20_interspeech,zhao23d_interspeech,xue2023contrast,zhang2024td}. We recall that, since no source code was available for these techniques, we use the scores reported in the papers, and we compare the performances only in terms of order of magnitudes. 

\paragraph{Comparison with supervised methods} We start with~\citep{kegler20_interspeech} and \citep{zhao23d_interspeech}, which are both based on deep supervised learning, as they break down PESQ and STOI performance per \textit{mask length} and thus allow a direct comparison with our results.
In~\citep{kegler20_interspeech}, with a training and test on a multi-speaker dataset (LibriSpeech) and in the informed inpainting case, the authors reported a PESQ (resp. STOI) score of \num{3.24}, \num{2.81}, and \num{2.18} (resp. \num{0.94}, \num{0.89}, and \num{0.73}) for mask lengths of \num{100}, \num{200}, and \SI{400}{\milli\second}, respectively. In~\citep{zhao23d_interspeech}, the authors reported PESQ (resp. STOI) scores of \num{3.30}, \num{2.61}, and \num{1.76} (resp. \num{0.96}, \num{0.89}, and \num{0.73}) for similar masks and dataset.  
In our study, the best scores on the same settings for \inpaintM~(resp. \inpaintI), were \num{3.13}, \num{2.93}, and \num{2.66} (resp. \num{3.06}, \num{2.70}, and \num{2.39}) for PESQ, and \num{0.93}, \num{0.88}, and \num{0.83} (resp. \num{0.90}, \num{0.85}, and \num{0.79}) for STOI (see \Cref{tab:inf}, multi-speaker).  

For the blind case, we can only compare our results to those reported in~\citep{kegler20_interspeech} (Table 3, condition ``FC-gaps") since it is not treated in~\citep{zhao23d_interspeech}, to the best of our understanding. In this case, our performances are significantly lower, both in terms of STOI and PESQ. For example, for a mask length of \SI{400}{\milli\second} \citep{kegler20_interspeech} reported a quite high STOI score of 0.71 when we obtained only 0.52 with the (best) framework \inpaintM. The differences between the two techniques are smaller when the mask length is shorter (e.g. a PESQ score of 2.72 in~\citep{kegler20_interspeech} for a mask length of \SI{200}{\milli\second}  vs. 2.31 with \inpaintM). Further experiments could be useful to better understand the origin of these differences in the case of blind inpainting. We would need to check that this is not simply due to the nature of the training/test datasets, to the analysis-synthesis ability of the methods, or to a different way of calculating the PESQ and STOI scores.

\paragraph{Comparison with self-supervised methods} We now have a look at~\citep{xue2023contrast} and \citep{zhang2024td}, as both methods used an SSL encoder, but fine-tuned to the task of inpainting. Comparison is more difficult with those studies, as experimental conditions are different from ours: they evaluate PLC, with several gaps per signal, while we evaluate mainstream inpainting with a single gap.
\citet{zhang2024td} reported a PESQ (resp. STOI) score of \num{3.27} and \num{2.35} (resp. \num{0.93} and \num{0.80}) for ``easy'' (\SI{15}{\percent} loss) and ``medium'' (\SI{30}{\percent} loss) conditions on a multi-speaker corpus (VCTK), in a blind setting. Our performance with \inpaintM is lower with both \SI{10}{\percent} loss (PESQ of \num{2.93} and STOI of \num{0.89}) compared to their easy condition, and \SI{40}{\percent} loss (PESQ of \num{1.53} and STOI of \num{0.52}) compared to their ``medium'' condition. \citet{xue2023contrast} reached an overall PESQ score of \num{2.88} with losses ranging from \SI{10}{\percent} to \SI{50}{\percent} in a blind setting, which is in the order of magnitude of our PESQ score with \SI{10}{\percent} loss (PESQ of \num{2.93}). Before drawing any conclusions, it should be reminded that the loss percentage reported by these studies accumulates the length of smaller gaps introduced into speech signals, and that the total length of those signals---accounting for the amount of uncorrupted speech---is not given by the authors. In our study, the loss corresponds to a single gap within a one-second signal. As a result, it is not clear which configuration is the most challenging, i.e., if the amount and position of uncorrupted portions of the input signal on which the SSL encoder relies is comparable between methods, and thus if those results could be compared with ours on an equal footing.
\newline

To conclude, this ``meta-comparison'' shows that the two proposed frameworks \inpaintI and \inpaintM seem to outperform other approaches based on supervised learning, at least in the informed case, and in particular for long masks (i.e., \SI{400}{\milli\second}). Here again, this can be explained by the ability of a powerful SSL model, pre-trained on a large amount of data, to extract the high-level linguistic information (e.g., syntactic and semantic) of the sentence to be reconstructed, based on the contextual non-missing information. 
Nevertheless, although the results should be treated with caution, it seems that when SSL encoders are introduced, the introduction of auxiliary modules and losses dedicated to the inpainting task are beneficial to model performance.

\begin{small}
\begin{table*}[t]
 \caption{Informed speech inpainting results with the \inpaintM framework on the out-of-domain test sets. For all metrics, average scores with confidence intervals for each test set and each mask length for both the single- and multi-speaker configurations.}

\begin{subtable}{\textwidth}
 \caption{Results on expressive speech (Expresso dataset).}
 \label{tab:inf_zeroshot}
 \vspace{0.25cm}
 \centering
 \begin{tabular}{c|c|ccc}
 \multicolumn{2}{c}{} & \multicolumn{3}{c}{\textbf{Multi-speaker (Expresso)}}\\
 \toprule
 Models & Mask (ms) & PESQ~$[-0.5;4.5]$ $\uparrow$ & STOI~$[0;1]$ $\uparrow$ & CER~(\%) $\downarrow$ \\ 
 \midrule 
 Unmasked   & 0 & $4.07 \pm 0.01$ & $0.96 \pm 0.01$ & $8 \pm 1$ \\
 \midrule 
        & 100 & $3.02 \pm 0.03$ & $0.91 \pm 0.01$ & $13 \pm 1$ \\
 \inpaintM  & 200 & $2.89 \pm 0.03$ & $0.84 \pm 0.01$ & $18 \pm 1$   \\
        & 400 & $2.64 \pm 0.03$ & $0.80 \pm 0.01$ & $24 \pm 2$ \\
 \bottomrule
 \end{tabular}
\end{subtable}
\vspace{0.5cm}

\begin{subtable}{\textwidth}
 \caption{Results on the noise-corrupted test sets. Results are broken down for each SNR and noise type.}
 \label{tab:inf_noisy}
 \vspace{0.25cm}
 \centering
 \begin{tabular}{c|c|c|cc||cc}
 \multicolumn{3}{c}{} & \multicolumn{2}{c}{\textbf{Single-speaker (LJ Speech)}} & \multicolumn{2}{c}{\textbf{Multi-speaker (VCTK)}}\\
 \toprule
  Noise type & Mask (ms) & SNR (dB) & PESQ~$[-0.5;4.5]$ $\uparrow$ & STOI~$[0;1]$ $\uparrow$ & PESQ~$[-0.5;4.5]$ $\uparrow$ & STOI~$[0;1]$ $\uparrow$\\ 
 \midrule 
    & & 20 & $3.20 \pm 0.32$ & $0.91 \pm 0.02$
    & $3.47 \pm 0.26$ & $0.96 \pm 0.02$ \\
    & 100 & 10 & $2.79 \pm 0.23$ & $0.89 \pm 0.02$
    & $3.18 \pm 0.25$ & $0.95 \pm 0.03$  \\
    & & 0 & $2.95 \pm 0.38$ & $0.85 \pm 0.05$ 
    & $2.98 \pm 0.30$ & $0.91 \pm 0.02$  \\ [0.05cm]
    \cline{2-7} 
    & & 20 & $2.84 \pm 0.28$ & $0.87 \pm 0.03$ 
    & $3.04 \pm 0.36$ & $0.94 \pm 0.03$  \\
White noise & 200  & 10 & $2.74 \pm 0.34$ & $0.81 \pm 0.05$ 
    & $3.03 \pm 0.38$ & $0.88 \pm 0.04$ \\
    & & 0 & $2.72 \pm 0.25$ & $0.85 \pm 0.06$ 
    & $2.80\pm 0.34$ & $0.87 \pm 0.05$ \\ [0.05cm]
    \cline{2-7}
    & & 20 & $2.75 \pm 0.26$ & $0.82 \pm 0.04$  
    & $2.88 \pm 0.34$ & $0.91 \pm 0.05$ \\
    & 400 & 10 & $2.43 \pm 0.27$ & $0.78 \pm 0.06$ 
    & $2.65 \pm 0.33$ & $0.87 \pm 0.06$ \\
    & & 0 & $2.27 \pm 0.30$ & $0.78 \pm 0.05$ 
    & $2.30 \pm 0.36$ & $0.79 \pm 0.06$ \\
 \midrule 
    & & 20 & $3.13 \pm 0.27$ & $0.91 \pm 0.02$ 
    & $3.35 \pm 0.29$ & $0.96 \pm 0.02$ \\
    & 100 & 10 & $3.01 \pm 0.28$ & $0.89 \pm 0.03$ 
    & $3.15 \pm 0.29$ & $0.94 \pm 0.02$ \\
    & & 0 & $2.82 \pm 0.31$ & $0.86 \pm 0.04$ 
    & $2.94 \pm 0.26$ & $0.91 \pm 0.03$ \\ [0.05cm]
    \cline{2-7}
    & & 20 & $2.96 \pm 0.30$ & $0.86 \pm 0.03$ 
    & $3.12 \pm 0.33$ & $0.94 \pm 0.03$ \\
Crowd noise & 200  & 10 & $2.74 \pm 0.31$ & $0.85 \pm 0.04$ 
    & $2.91 \pm 0.32$ & $0.91 \pm 0.03$ \\
    & & 0 & $2.58 \pm 0.29$ & $0.84 \pm 0.06$ 
    & $2.71 \pm 0.32$ & $0.84 \pm 0.04$ \\ [0.05cm]
    \cline{2-7}
    & & 20 & $2.81 \pm 0.27$ & $0.84 \pm 0.04$ 
    & $2.88 \pm 0.35$ & $0.90 \pm 0.04$ \\
    & 400 & 10 & $2.50 \pm 0.33$ & $0.82 \pm 0.05$ 
    & $2.75 \pm 0.32$ & $0.87 \pm 0.05$ \\
    & & 0 & $2.31 \pm 0.32$ & $0.79 \pm 0.05$ 
    & $2.51 \pm 0.34$ & $0.80 \pm 0.06$ \\
\bottomrule
\end{tabular}
\end{subtable}

\end{table*}
\end{small}

\subsection{Extrapolation to other datasets capabilities}

We now explore the extrapolation capabilities of our pre-trained framework \inpaintM to out-of-domain datasets, in the informed case. We omitted here the \inpaintLI and \inpaintT baselines which presented the lowest scores in previous evaluations.

\paragraph{Evaluation on multi-speaker expressive speech}

\Cref{tab:inf_zeroshot} reports the results of the objective evaluation in terms of PESQ, STOI and CER scores.
We assessed the significance of \textit{mask length} (\SI{100}, \SI{200}, \SI{400}{\milli\second}) with \textit{utterance} as a random factor. 
Statistical analysis shows that \textit{mask length} had a significant effect on each objective metric, and the \textit{utterance} random factor had no significant effect on PESQ and was therefore removed from this model.
Pairwise comparisons show significant differences between the three \textit{mask length} levels, on all metrics, corresponding to a 
decrease in performance with increasing mask lengths.

Although we observe a drop in all scores when tested on the expressive speech data compared to the evaluation on the in-domain VCTK test set (\Cref{tab:inf}), \inpaintM still achieves higher scores than work reported in other studies~\citep{kegler20_interspeech,zhao23d_interspeech} which only performed in-domain testing (see \Cref{sec:otherstudies}). 
Therefore, this demonstrates the robustness of our framework to zero-shot adaptation to both unseen speakers and unseen speaking style in the particular case of expressive speech synthesis.

\paragraph{Evaluation on noise-corrupted speech}

\Cref{tab:inf_noisy} reports the results of the objective evaluation of \inpaintM in terms of PESQ and STOI scores for white noise and crowd noise corruption, respectively.
We assessed here the significance of \textit{mask length} (\SI{100}, \SI{200}, \SI{400}{\milli\second}), \textit{SNR} (\SI{10}, \SI{20}, \SI{40}{\dB}), \textit{noise type} (white noise and crowd noise) and \textit{dataset} (single- and multi-speaker) with their interactions.

Statistical analysis of PESQ distributions 
showed that the four-way interaction (\textit{mask length} $\times$ \textit{SNR} $\times$ \textit{noise} $\times$ \textit{dataset}) as well as all three-way and two-way interactions were not significant. This allowed to simplify the statistical model by removing one interaction at a time while assessing if the removal of each interaction had a significant impact on the model. Ultimately, only the \textit{mask length} and \textit{SNR} were found to have a significant effect on the PESQ distributions. 
The effect \textit{mask length} shows the same trends as in clean speech, with a decrease of performance with longer masks. Then, the global decrease of PESQ across all conditions with decreasing \textit{SNR} displays a sensitivity to the amount of corruption noise that is introduced. 
Interestingly, the non-significance of the \textit{noise type} factor indicates that the \inpaintM method is equally robust to either type of noise. Similarly,
the absence of a \textit{dataset} effect indicates that the addition of noise has
levelled scores to a point that mitigate the impact of the training corpus on the model.

Statistical analysis of STOI distributions 
also gave a simple model, where only the \textit{mask length} and the \textit{SNR} $\times$ \textit{dataset} interaction have a significant effect. We find again a decreasing of STOI with longer masks.
As for the interaction, multiple comparisons show that the effect of \textit{SNR} is only present in the multi-speaker condition, while
intelligibility is little affected by the amount of corruption noise in the single-speaker condition.
Compared to PESQ, we found a significant effect of \textit{datasets} for SNRs of \SI{10} and \SI{20}{\dB} only. Therefore, the differences in intelligibility scores observed on clean speech between the single-speaker and the multi-speaker \textit{datasets} remain up to the highest level of corruption noise.
The absence of effect of the \textit{noise type} factor again shows that all observations hold for both white noise and crowd noise.

To conclude, we observed very similar PESQ and STOI scores between white noise- and crowd noise-corrupted test sets, supported by the non significance of the \textit{noise type} factor. Moreover, it seems that the addition of corruption noise has affected both PESQ and STOI performances to the point where most other factors have less impact on the objective measures. 
In the end, \inpaintM 
PESQ scores are even comparable with those obtained on clean speech for both corruption noises (\Cref{tab:inf}).
This proves a high robustness of this model to corrupted input signal.
Last but not least, it is important to note that the inpainting of noise-corrupted signals generates a portion of signal which is without noise.
Thus, while this experiment is valuable to demonstrate the robustness of our inpainting method to the corruption of the signal used as context, the resulting output signal is not currently relevant for a direct application to a speech generation use case.

\section{Conclusion}
\label{section:conclusion}

This study investigated the extent to which contextual speech representations extracted by an SSL model can effectively drive an inpainting process, i.e., reconstructing a missing portion of a speech signal from its surrounding context. More specifically, our hypothesis was that it is possible to perform speech inpainting with an SSL encoder that is pre-trained on an unmasking task, only by appending a decoder to convert back to the time domain.
We compared two different frameworks for this sake. First, a pre-trained HuBERT SSL encoder with an adapted HiFi-GAN decoder (\inpaintM) was used to test our hypothesis. Second, a fine-tuned HuBERT SSL encoder to the inpainting task with a pre-trained vanilla HiFi-GAN (\inpaintI) was used as a reference to state-of-the-art methods that all train their model specifically for the inpainting task.
Additionally, we explored an alternative approach inspired by text-informed inpainting, which combines ASR with zero-shot TTS (\inpaintT). We evaluated our main method on both in-domain and out-of-domain datasets to evaluate its robustness to a variety of input signals.

As a main result, we validated our main hypothesis, by demonstrating that a pre-trained SSL-based approach (\inpaintM) can successfully perform inpainting tasks both in terms of quality and intelligibility, as measured with both objective and perceptive metrics. \inpaintM systematically outperformed our baselines, including \inpaintT which performed inpainting using a text-based language model. Interestingly, when compared to other studies in the literature, \inpaintM was extremely competitive with the most recent supervised methods, while our method was not specifically trained for the inpainting task, but indirectly trained on unmasking. This clearly demonstrates the potential of SSL encoders to transfer knowledge between unmasking and inpainting without the need for fine-tuning, provided that a decoder is appended to generate a speech waveform. Moreover, we showed that such method is resilient to varied input data including expressive speech and noise-corrupted speech---two challenging conditions that, to the best of our knowledge, have received limited attention in previous inpainting studies. 

Then, we evaluated the benefit of fine-tuning the SSL encoder to the inpainting task. We observed a symmetry in our results, with fine-tuning the SSL encoder (\inpaintI) being the most effective strategy for inpainting in a single-speaker setting, whereas the pre-trained encoder (\inpaintM) yielded better results in a multi-speaker scenario. The higher performance of \inpaintI on the single-speaker supports the hypothesis that while a pre-trained encoder is already successful in inpainting, the addition of further supervised training is beneficial to the task. In the multi-speaker setting, one main difference between models is the absence of a speaker embedding when fine-tuning the SSL encoder (\inpaintI). This suggests that the HuBERT SSL encoder may not keep enough speaker information, even in a supervised inpainting learning task, to reconstruct the signal with similar quality than in the pre-trained configuration. 
Alternate solutions for fine-tuning that put the SSL encoder in a more complex architecture comprising auxiliary modules and varied training losses, such as \citep{xue2023contrast,zhang2024td}, seem to be relevant when looking at performance scores reported in those paper. However, their experimental set-ups are not fully comparable with ours.

Finally, while specifically fine-tuning models on inpainting tasks in a supervised fashion is and will certainly remain the most efficient way to increase performance, we took a step aside in this paper. 
We followed a more hypothesis-driven approach based on a controlled and symmetric experimental protocol (pre-trained vs. fine-tuned SSL) to quantitatively assess the relationships between the pretext task of SSL (unmasking) and the inpainting downstream task. 
By demonstrating successful transfer between the two tasks, this paves the way for further use of this simple single SSL encoder configuration for inpainting, as a tool for understanding the speech inpainting process and its relationship to language and cognition. In particular, future
work may focus on (i) benchmarking other SSL models for inpainting, including models not fine-tuned for ASR, as well as alternatives such as wav2vec \citep{Baevski2020}, and WavLM \citep{Chen2022}; (ii) a fine-grained analysis of the inpainted speech at different linguistic scales (phonetic, syllabic, morphologic); (iii) performing an in-depth study of the impact of the codebook size $\mathcal{C}$ on overall performance; and (iv) the relationship between the context actually used by the SSL encoder on one hand, and the length and linguistic complexity of the signal to be reconstructed, on the other. 
Finally, in addition to their technological applications, the proposed speech inpainting systems, and SSL models in general, provide a means of finely quantifying the amount of predictable information in the speech signal. Therefore, they can  be potentially useful for studying, through computational modelling and simulation, some of the predictive  processes underlying  speech perception \citep{friston2009predictive,tavano2015prediction}. The proposed framework based on non-causal prediction could complement other studies conducted in the context of the predictive coding framework and focusing on causal predictions (e.g., \citep{10.1162/neco_a_01264,caucheteux2023evidence, Heilbron2019TrackingNL}).

\section*{Acknowledgment}

This work has been partially supported by MIAI Cluster (ANR-23-IACL-0006), MIAI@Grenoble Alpes, (ANR-19-P3IA-0003), and by ANRT CIFRE.

\section*{CRediT authorship contribution statement}

\textbf{Ihab Asaad:} Conceptualization, Methodology, Software, Validation, Investigation, Data curation, Writing – original draft. 
\textbf{Maxime Jacquelin:} Conceptualization, Methodology, Software, Validation, Investigation, Data curation, Writing – original draft. 
\textbf{Olivier Perrotin:} Conceptualization, Methodology, Formal analysis, Writing – original draft, Writing – review \& editing, Visualization, Supervision, Funding acquisition.
\textbf{Laurent Girin:} Conceptualization, Formal analysis, Writing – original draft, Writing – review \& editing, Supervision, Funding acquisition.
\textbf{Thomas Hueber:} Conceptualization, Methodology, Software, Validation, Investigation, Resources, Writing – original draft, Writing – review \& editing, Supervision, Funding acquisition, Project administration.

\section*{Declaration of generative AI and AI-assisted technologies in the writing process}

During the preparation of this work the author(s) used \textit{DeepL} and \textit{Le Chat - Mistral AI} in order to polish the English of some paragraphs. After using this tool/service, the author(s) reviewed and edited the content as needed and take(s) full responsibility for the content of the publication.

\bibliographystyle{elsarticle-harv} 
\bibliography{mybib}






\end{document}